\documentclass[12pt]{article}
\newcommand{\quattrova}{($\varphi^{\scriptscriptstyle a},
c^{\scriptscriptstyle a},
\lambda_{\scriptscriptstyle  a},{\o c}_{\scriptscriptstyle a}$)~}

\newcommand{\be}{\begin{equation}}
\newcommand{\ee}{\end{equation}}
\newcommand{\bea}{\begin{eqnarray}}
\newcommand{\eea}{\end{eqnarray}}
\def \Qb{ Q_{\scriptscriptstyle BRS}}
\def \QBb{{\overline {Q}}_{\scriptscriptstyle BRS}}
\def \HT{{\widetilde{\mathcal H}}}
\def \LT{{\widetilde{\mathcal L}}}

\def \QH{{Q_{\scriptscriptstyle H}}}
\def \e{\epsilon}
\def \o{\overline}
\def \l{\lambda}
\def \bc{\o{c}}
\def \w{\omega}
\def \s{\scriptscriptstyle}
\def \p{\varphi}
\def \QBH{{\overline{Q}}_{\scriptscriptstyle H}}
\def \Qb{ Q_{\scriptscriptstyle BRS}}
\def \QBb{{\overline {Q}}_{\scriptscriptstyle BRS}}
\def \NH{ N_{\scriptscriptstyle H}}
\def \NHB{\overline \NH}

\newcommand{\scite}{~\cite}

\def \LTloc{\LT_{susy}}

\def \bpsi{\o{\psi}}
\def \pipsi{\Pi_{\scriptscriptstyle \psi}}
\def \pipsibar{\Pi_{\scriptscriptstyle\o \psi}}
\def \pigi{\Pi_{\scriptscriptstyle g}}
\def \gaug{(\psi,{\o\psi},g)}
\def \HTloc{\HT_{susy}}

\begin{document}

\baselineskip =15.5pt
\pagestyle{plain}
\setcounter{page}{1}

\begin{flushright}
\end{flushright}
\vfil

\begin{center}
{\huge On the ``Universal" N=2 Supersymmetry of Classical Mechanics.}
\end{center}

\vfil

\begin{center}
{\large E.Deotto, E.Gozzi}\\
\vspace {1mm}
Dipartimento di Fisica Teorica, Universit\`a di Trieste, \\
Strada Costiera 11, P.O.Box 586, Trieste, Italy \\ and INFN, Sezione 
di Trieste.\\
\vspace {1mm}
\vspace{3mm}
\end{center}

\vfil

\begin{abstract}

\noindent
In this paper we continue the study of the geometrical features
of a functional approach to classical mechanics proposed some time ago.
In particular we try
to shed some light on a N=2 ``universal" supersymmetry which seems to have an
interesting interplay with the concept of ergodicity of the system.
To study the geometry better we make this susy local and clarify
pedagogically several issues present in the literature. Secondly,
in order to prepare the ground for a better understanding of its relation
to ergodicity, we study the system on constant energy surfaces. We
find that the procedure of constraining the system on
these surfaces injects in it some local Grassmannian invariances
and reduces the N=2 global susy to an N=1.
\end{abstract}

\section {Introduction}
\indent
In this paper we continue the study of a path-integral approach
to Classical Mechanics (CM) started in ref.\scite{Ennio}. This path-integral
is nothing else than the functional counterpart of an {\it operatorial}
approach to CM pioneered by Koopman and von Neumann\scite{Koop} in the 30's.
This operatorial formulation is basically the one where the evolution
is given by the Liouville operator.

In our approach we had a lot of extra variables besides those labelling
the phase-space of the original mechanical system. It was discovered
\scite{Ennio}\scite{Geom} that these extra variables
had a beautiful  geometrical meaning. They were basically
the basis for the forms and the tensor fields which 
one could build out
of the tangent and cotangent bundles to phase-space.
As a consequence our approach  not only reproduced the Liouville operator
of Koopman-von Neumann, which in geometrical
terms\scite{Marsd} is the Hamiltonian vector field associated to the
motion, but it generated automatically the entire Lie-derivative 
which is necessary for the motion of higher forms and tensors.

These extra variables were anyhow redundant and this redundancy
was signalled by the presence of some ``universal" symmetries
which made a superalgebra known as Isp(2). The meaning of the
charges associated to these symmetries was also intrinsically
geometrical. In fact they turned out  to be related to the exterior
derivatives on phase-space and the form-number operator. The
entire Cartan calculus on phase-space could be reproduced via our
``universal" charges. All this will be briefly summarized
in section 2 of this paper.

Beside these symmetries, it was later
discovered\scite{Ergo} a new one which was actually a
non-relativistic supersymmetry (susy). Those authors, anyhow,
did not manage to understand the geometrical meaning of that susy 
and this paper
is an attempt in that direction. 
That susy had\scite{Ergo} also a nice 
interplay
with the concept of ergodicity\scite{Avez} of the dynamical system
under study and we think that it is
crucial to get a better understanding of  this interplay.
  
To tackle the first issue, i.e. the geometrical aspects of our susy,
the direction we take
is to make the susy local and study in detail what we
obtain. We say ``study in detail" because in the literature
there were some strange statements\scite{Alv} claiming to show that,
at least for the supersymmetric QM of Witten\scite{Witte}, the lagrangian
with local-susy was equivalent to the one with global susy. We shall
show that it is not so. One should actually perform very carefully the full
Dirac\scite{Vinc} procedure or, via path-integrals, 
the Faddeev procedure\scite{Pavao}
or apply the BFV methodology \scite{Teitel} of handling systems
with constraints. If one does that carefully it is easy to realize that the
system with local susy has a different number of degrees of
freedom than  the one with global susy. The states themselves are
restricted to the so called physical states by the presence of the 
local symmetry. It was this last step that was missing in ref.\scite{Alv}
and which led to the wrong conclusion. We show all this in great details
in section 3 of this paper.

The physical states condition and the BFV procedure
are  what 
lead us, in section 4 of the paper, to understand the geometrical 
meaning of the susy charges.
They turn out to be an essential ingredient to
restrict the forms to the so called equivariant ones
\scite{Cart}\scite{Duit}\scite{Berline}. The business of 
equivariant cohomology has popped out
recently in the literature in connection with  topological field theories
\scite{Topol}. Some attempts\scite{Niem} had been done in the past of
cooking up a BRS-BFV charge which would produce as physical states
the equivariant ones but without showing from which local symmetry
this BRS-BFV charge was coming from. Here we have filled that gap that means 
we have shown in details which local symmetry gives rise to a BFV charge 
whose physical states are the equivariant ones.   We think that
providing, as we have done here, all the details of the path-integral 
and local symmetry construction will be helpful to the readers 
even for just purely pedagogical  reasons.

In  section 5 of the paper we turn to the other aspect
of our supersymmetry, that is its interplay with the concept of
ergodicity\scite{Ergo}. To get a better grasp of this
problem we realized long ago\scite{Ergo} that we had to formulate our
functional approach on constant energy surfaces.
So here somehow we reverse the procedure followed in the
previous sections. There we basically  had found that the local supersymmetries
constrain the states and the motion to some hypersurfaces of our enlarged
space, here instead we shall constrain by hand the system to move on some 
fixed hypersurfaces,
the constant energy ones, and check what happens.
We realize that in this formulation the energy
plays the role of a coupling and it turns out to be associated to a 
tadpole term of the new lagrangian. Moreover  we find that by
constraining the system on constant energy surfaces we gain  a {\it local} 
graded symmetry which is not anyhow a local susy. Regarding
the  {\it global} symmetries we lose part of the original global N=2 susy 
which is  now reduced  to an N=1. This fact, which may appear as a bad feature of the
procedure, may actually  turn out to be a  virtue. In fact we shall  have
to study the interplay of ergodicity and susy by means of only one susy charge
and not two as before. Anyhow the detailed study of this interplay  
will be left to future papers where we will concentrate more on dynamical
issues and not on geometrical ones as we have done here. Some detailed 
calculations are confined to few appendices.
 
\section {Review of the Functional Approach To Classical Mechanics.}
\indent
We will briefly review here what is contained in ref.\scite{Ennio}.
In those papers the authors gave a {\it path integral} formulation to CM.
This may sound absurd but we should remember that whenever a theory
has an {\it operatorial} formulation it has also a {\it path-integral} one.
Now CM  has an operatorial formulation proposed long ago by Koopman 
and von-Neumann\scite{Koop}. This operatorial approach is basically
the one where the time evolution of phase-space distributions
is governed by the~{\it Liouville}  operator~ or for higher forms by 
the {\it Lie derivative}  of the Hamiltonian flow\scite{Marsd}.
So, if there is an operatorial approach, there should also be a
path-integral one which we will indicate from now on as CPI for 
{\bf C}lassical {\bf P}ath {\bf I}ntegral.

In CM we  have a  $2n$-dimensional phase space ${\cal M}$ 
whose coordinates we call\break $\varphi^{a}$$(a=1,\ldots 2n)$, i.e.: 
$\varphi^{a}=(q^1\cdots  q^n,p^1\cdots p^n)$ and we indicate the  
Hamiltonian of the system as $H(\varphi)$ while the symplectic matrix is  
$\omega^{ab}$.
The equations of motion are then:

\begin{equation}
\label{eq:uno}
{\dot\varphi }^{a}=\omega^{ab}{\partial H\over\partial\varphi ^{b}} \label{uno}
\end{equation} \vspace{0 mm} $\!\!\!$

We shall now introduce the {\it classical} analog,~
$Z_{\scriptscriptstyle CM}$,  of the quantum generating 
functional:

\begin{equation}
\label{eq:due}
Z_{\scriptscriptstyle CM}[J]=N\int~{\cal D}\varphi ~{\tilde{\delta}}[\varphi 
(t)-\varphi _{cl}(t)]~\exp\int 
~J\varphi \;dt \label{due}
\end{equation} \vspace{0 mm} 

\noindent where $\varphi$ are the $\varphi^{a}\in{\cal M}$, $\varphi_{cl}$ are 
the solutions of eq.(\ref{uno}),
$J$ is an external current and $\widetilde{\delta}[\;\;]$ is a functional 
Dirac-delta which forces
every path $\varphi(t)$ to sit on a classical one $\varphi_{cl}(t)$.  
There are all
possible initial conditions integrated over in (\ref{due}) and, because of this,
one should be very careful in properly defining the measure 
of integration and the functional
Dirac delta. 

We should now check whether the path integral of eq.~(\ref{due}) leads to the well 
known operatorial 
formulation\scite{Koop} of CM done via the Liouville operator and the Lie 
derivative. To do that let us first
rewrite the functional Dirac delta in (\ref{due}) as:

\begin{equation}
\label{eq:tre}
{\tilde\delta}[\varphi -\varphi _{cl}]={\tilde\delta}[{\dot\varphi 
^{a}-\omega^{ab}
\partial_{b}H]~\det [\delta^{a}_{b}\partial_{t}-\omega^{ac}\partial_{c}\partial
_{b}H}]
\end{equation} \vspace{0 mm} 

\noindent where we have used the analog of the relation
$\delta[f(x)]=\frac{\delta[x-x_i]}{\Bigm|\frac{\partial f}{\partial 
x}\Bigm|_{x_i}}$. 

\noindent The determinant which appears in (\ref{eq:tre}) is always positive and
so we can drop the modulus sign $|\;\;|$. 
The next step is to insert (\ref{eq:tre}) in (\ref{due}) and write the 
$\tilde{\delta}[\;\;]$
as a Fourier transform over some new variables $\lambda_{a}$, i.e.:

\begin{equation}
\label{eq:quattro}
{\tilde{\delta}}\biggl[{\dot\varphi }^{a}-\omega^{ab}{\partial 
H\over\partial\varphi ^{b}}\biggr]=
\int~{\cal D}\lambda_{a}~\exp~i\int\lambda_{a}\biggl[{\dot \varphi 
}^{a}-\omega^{ab}
{\partial H\over\partial\varphi ^{b}}\biggr]dt\nonumber
\end{equation} \vspace{0 mm} 
Next we express the determinant 
$\det[\delta^{a}_b\partial_t-\omega^{ac}\partial_c\partial_bH]$ via
Grassmannian variables $\o{c}_a, c^{a}$:

\begin{equation}
\label{eq:cinque}
\det[\delta^{a}_{b}\partial_{t}-\omega^{ac}\partial_{c}\partial_{b}H]
=\int~{\cal D}c^{a}{\cal D}{\o c}_{a}~\exp~\biggl[-\int {\o c}_{a}
[\delta^{a}_{b}
\partial_{t}-\omega^{ac}\partial_{c}\partial_{b}H]c^{b}~dt\biggr]
\end{equation} \vspace{0 mm} 

\noindent Inserting (\ref{eq:quattro}), (\ref{eq:cinque}) and (\ref{eq:tre}) 
in (\ref{due}) we get:

\begin{equation}
\label{eq:sei}
Z_{\scriptscriptstyle CM}[0]=\int~{\cal D}\varphi ^{a}{\cal D}\lambda_{a}{\cal 
D}c^{a}{\cal D}
{\o c}_{a}~\exp~\biggl[i\int~dt~{\widetilde{\cal L}}\biggr]
\end{equation} \vspace{0 mm} 

\noindent where $\widetilde{\cal L}$ is:

\begin{equation}
\label{eq:sette}
{\widetilde{\cal L}}=\lambda_{a}[{\dot\varphi }^{a}-\omega^{ab}\partial_{b}H]+
i{\o c}_{a}[\delta^{a}_{b}\partial_{t}-\omega^{ac}\partial_{c}\partial_{b}H]
c^{b}
\end{equation} \vspace{0 mm} 
The variation of the action associated to this lagrangian gives the following equations 
of motion:
\begin{eqnarray}
\label{eq:otto}
{\dot\varphi }^{a}-\omega^{ab}\partial_{b}H & = & 0 \\
\label{eq:nove}
[\delta^{a}_{b}\partial_{t}-\omega^{ac}\partial_{c}\partial_{b}H]c^{b}
 & = & 0 \\
\label{eq:dieci}
\delta^{a}_{b}\partial_{t}{\o c}_{a}+{\o
c}_{a}\omega^{ac}\partial_{c}\partial_{b}H & = & 0 \\
\label{eq:undici}
[\delta_{b}^{a}\partial_{t}+\omega^{ac}\partial_{c}\partial_{b}H]\lambda_{a}
& = & -i{\o c}_{a}\omega^{ac}\partial_{c}\partial_{d}\partial_{b}H c^{d}
\end{eqnarray} \vspace{0 mm} 
\noindent
From these equations we gather that~$\widetilde{\cal L}$ leads to the same Hamiltonian equations for 
$\varphi$ as~$H$~did and that $c^b$ transforms under the Hamiltonian vector field
\scite{Marsd} $h\equiv\omega^{ab}
\partial_bH\partial_{a}$ as a {\it form} $d\varphi^{b}$ does.

The Hamiltonian $\widetilde{\cal H}$ associated to the lagrangian (7) is:

\begin{equation}
\label{eq:dodici}
\widetilde{\cal 
H}=\lambda_a\omega^{ab}\partial_bH+i\o{c}_a\omega^{ac}
(\partial_c\partial_bH)c^{b}
\end{equation} 

\noindent and via some
super-extended Poisson brackets ({\it EPB}) defined in the space
\quattrova\break
one can re-obtain the equations of motion (8)-(11). If we had considered
all  the variables
\quattrova as configurational ones, then  there would have been  constraints 
among these variables and
the associated momenta. In that case we would have had to adopt 
the Dirac procedure
\scite{Vinc}. It will be  explained in ref.\scite{Hilb} how this 
can be done and that
the {\it EPB}-brackets above will  be exactly the brackets 
produced by the Dirac method.
These extended Poisson brackets are:
\begin{equation}
\label{eq:tredici}
\{\varphi ^{a},\lambda_{b}\}_{\scriptscriptstyle EPB}
=\delta^{a}_{b}~~;~~\{{\o c}_{b},
c^{a}\}_{\scriptscriptstyle EPB}=-i\delta^{a}_{b} 
\end{equation} \vspace{0 mm} 
All the other {\it EPB} are zero, in particular $\{\varphi^{a},\varphi^{b}\}_
{\scriptscriptstyle EPB}=0$. Note from this relation
that the {\it EPB}  are not the standard Poisson
brackets on ${\cal M}$ which would have given: $\{\varphi^{a},\varphi^{b}\}_
{\scriptscriptstyle PB}=\omega^{ab}$.

Since (\ref{eq:sei}) is a path integral, one could also introduce the concept of 
{\it commutator}
as Feynman did in the quantum case. If we define the graded commutator of two 
functions $O_1(t)$ and
$O_2(t)$ as the expectation value 
$\langle\;\;\;\rangle$ under our path integral of some time-splitting 
combinations of the functions
themselves, as:

\begin{equation}
\label{eq:quattordici}
\langle[O_{1}(t),O_{2}(t)]\rangle\equiv  \lim_{\epsilon\rightarrow 0}
\langle O_{1}(t+\epsilon)O_{2}(t)\pm O_{2}(t+\epsilon)O_{1}(t)\rangle 
\end{equation} \vspace{0 mm} 

\noindent then we get from (\ref{eq:sei}) that the only commutators  
different from zero are:

\begin{equation}
\label{eq:quindici}
\langle[\varphi ^{a},\lambda_{b}]\rangle=i\delta^{a}_{b}~~;~~\langle[{\o 
c}_{b},
c^{a}]\rangle=\delta^{a}_{b}.
\end{equation} \vspace{0 mm} 

\noindent Note that there is an isomorphism between the extended Poisson structure 
(\ref{eq:tredici}) and the graded commutator structure (\ref{eq:quindici}):
\begin{equation}
\label{eq:sedici}
\{\cdot,\cdot\}_{\scriptscriptstyle EPB}\longrightarrow -i[\cdot,\cdot]
\end{equation}
and we will always use  the second one from now on.
The commutator structure (\ref{eq:quindici}) allow us to ``realize" 
$\lambda_a$ and $\o{c}_a$ as:

\begin{equation}
\label{eq:diciassette}
\lambda_{a}=-i{\partial\over\partial\varphi ^{a}}~~;~~{\o 
c}_{a}={\partial\over
\partial c^{a}}
\end{equation}  

\noindent Now we have all the tools to turn  the weight in (\ref{eq:sei})
 into an operator. For the moment let us take only
the bosonic part of $\widetilde{\cal H}$:

\begin{equation}
\label{eq:diciotto}
{\widetilde{\cal H}}_{\scriptscriptstyle B}=\lambda_{a}\omega^{ab}\partial_{b}H
\end{equation} 

\noindent This one, via (\ref{eq:diciassette}), turns into the operator:

\begin{equation}
\label{eq:diciannove}
{\widehat{\widetilde{\cal H}}}_{\scriptscriptstyle B}\equiv 
-i\omega^{ab}\partial_{b}H\partial_{a}
\end{equation}  

\noindent which is the Liouville operator of CM. If we had added the 
Grassmannian part to 
${\widetilde{\cal H}}_{\scriptscriptstyle B}$ and inserted the operatorial 
representation (\ref{eq:diciassette}) of 
$\o{c}$ , we would have got the Lie-derivative of the
Hamiltonian flow as we shall show later .
So this proves that the operatorial version of CM comes from  
a path-integral weight which is  just a Dirac delta on 
the classical paths. Somehow this is the {\it classical}
analogue of what Feynman did for {\it Quantum}
Mechanics where he proved that the Schroedinger operator of evolution
comes from a path-integral weight of the form $\exp\,(i S)$.

We have seen before that $c^{a}$ transform as  $d\varphi^{a}$, that is as 
the {\it basis} of generic forms $\alpha\equiv \alpha_{a}(\varphi)d\varphi^{a}$, 
but they also transform   as the
{\it components} of  tangent vectors:
$V^{a}(\varphi){\partial\over\partial\varphi^{a}}$.
The space whose coordinates are $(\varphi^{a}, c^{a})$ is called, in 
ref.\scite{Schw}, 
the {\it reverse-parity} tangent bundle and it is indicated as $\Pi T{\cal M}$. The 
``{\it reversed-parity}" specification
is because the $c^{a}$ are Grassmannian variables. As the $(\lambda_{a}, 
{\o c}_{a})$ are the ``momenta" of the previous 
variables (see eq.(\ref{eq:sette})) we conclude that the $8n$ variables \quattrova~span the cotangent 
bundle to the
reversed-parity tangent bundle:~$T^{\star}(\Pi T{\cal M})$.
So our super-space is a cotangent bundle and this is the reason why it has a 
Poisson structure which is the one we found via the CPI and indicated in
eq. (\ref{eq:tredici}). For more details about this we refer the interested
reader to ref.\scite{Geom}.

In the remaining part of this section we will show how to reproduce all the 
abstract Cartan calculus
via our commutators and the Grassmannian variables. Let us first introduce five 
charges which are
conserved under the $\widetilde{\cal H}$ of eq. (\ref{eq:dodici}) and which 
will play an important role in the Cartan calculus. They are:
\begin{eqnarray}
\label{eq:venti}
&& Q_{\scriptscriptstyle BRS} \equiv i c^{a}\lambda_{a} \\
\label{eq:ventuno}
&& {\o Q}_{\scriptscriptstyle BRS} \equiv i {\o
c}_{a}\omega^{ab}\lambda_{b} \\
\label{eq:ventidue}
&& Q_{g} \equiv c^{a}{\o c}_{a} \\
\label{eq:ventitre}
&& K\equiv {1\over 2}\omega_{ab}c^{a}c^{b} \\
\label {eq:ventiquattro}
&& {\o K} \equiv {1\over 2}\omega^{ab}{\o c}_{a}{\o c}_{b}
\end{eqnarray}  
\noindent The $\omega_{ab}$ are the matrix elements of the inverse of 
$\omega^{ab}$. These five charges make
a superalgebra which we \scite{Ennio} called Isp(2) for inhomogeneous symplectic
group. The reason for the name will be clear if we use a superspace
as it was done in ref.\scite{Ennio}.

Now since $c^{a}$ transforms  under the Hamiltonian flow
as the basis $d\varphi^{a}$ of forms and  $\o{c}_a$  transforms  as the basis 
of  vector fields\footnote{Note that $\lambda_{a}$
does not seem to transform as a vector field, eq. (\ref{eq:undici}), even if it 
can be  interpreted as $\frac{\partial}{\partial \varphi^{a}}$. The
explanation of this fact is given in  the second paper of ref.~\scite{Geom}.},
(see eq. (\ref{eq:dieci})), let us start building the following map, 
called ``hat" map $\wedge$:
\begin{eqnarray}
\label{eq:venticinque}
\alpha=\alpha_{a}d\varphi ^{a} & \hat{\longrightarrow} &  {\widehat\alpha}\equiv
\alpha_{a}c^{a}\\
\label{eq:ventisei}
V=V^{a}\partial_{a}  & \hat{\longrightarrow} & {\widehat V}\equiv V^{a}{\o
c}_{a}
\end{eqnarray}  
\noindent It is  actually a much more general map between forms $\alpha$, 
antisymmetric
tensors $V$ 
and functions of $\varphi, c, \o{c}$:

\begin{eqnarray}
\label{eq:ventisette}
F^{(p)}={1\over p !}F_{a_{1}\cdots a_{p}}d\varphi ^{a_{1}}\wedge\cdots\wedge
d\varphi ^{a_{p}}  & \hat{\longrightarrow} &{\widehat F}^{(p)}\equiv {1\over p!}
F_{a_{1}\cdots a_{p}}c^{a_{1}}\cdots c^{a_{p}}\\
\label{eq:ventotto}
V^{(p)}={1\over p!}V^{a_{1}\cdots a_{p}}\partial_{a_{1}}\wedge\cdots\wedge 
\partial_{a_{p}} & \hat{\longrightarrow} & {\widehat V}\equiv {1\over 
p!}V^{a_{1}
\cdots a_{p}}{\o c}_{a_{1}}\cdots {\o c}_{a_{p}}
\end{eqnarray} 

\noindent Once the correspondence (\ref{eq:venticinque})-(\ref{eq:ventotto}) is 
extablished,  we can easily find  out what corresponds to the various Cartan operations 
like  the exterior derivative ~$d$~
of a form, the  interior contraction $\iota_{{\scriptscriptstyle V}}$
between a vector field $V$ and a form $F$ and
the multiplication of a form by its form number \scite{Ennio}  :
\begin{eqnarray}
\label{eq:ventinove}
dF^{(p)} & \hat{\longrightarrow} & [Q_{\scriptscriptstyle BRS},{\widehat 
F}^{(p)}] \\
\label{eq:trenta}
\iota_{{\scriptscriptstyle V}}F^{(p)} & \hat{\longrightarrow} & [{\widehat V},
{\widehat F}^{(p)}]
\\
\label{eq:trentuno}
pF^{(p)} & \hat{\longrightarrow} & [Q_{g}, {\widehat 
F}^{(p)}]
\end{eqnarray} 
\noindent where $Q_{\scriptscriptstyle BRS}, \, Q_g$ are the charges of 
(\ref{eq:venti})-(\ref{eq:ventidue}).
In the same manner we can translate in our language the usual 
mapping\scite{Marsd}
between vector fields $V$
and forms $V^{\flat}$ realized by the symplectic 2-form $\omega(V,0)\equiv 
V^{\flat}$,
or the inverse operation of building a vector field $\alpha^{\sharp}$ out of a 
form:
$\alpha=(\alpha^{\sharp})^{\flat}$. These operations can be translated in our
formalism as follows:
\begin{eqnarray}
\label{eq:trentadue}
V^{\flat} & \hat{\longrightarrow} & [K,{\widehat V}]\\
\label{eq:trentatre}
\alpha^{\sharp} & \hat{\longrightarrow} & [{\o
K},{\widehat\alpha}]
\end{eqnarray}  
\noindent where again $K, \o{K}$ are the charges 
(\ref{eq:ventitre})-(\ref{eq:ventiquattro}).
We can also translate the standard operation of building an
Hamiltonian vector field, indicated as $(df)^{\sharp}$, out of a
function ~$f(\varphi)$, and also the Poisson brackets between two 
functions $f$ and $g$:
\begin{eqnarray}
\label{eq:trentaquattro}
(df)^{\sharp} & \hat{\longrightarrow} & [{\o Q}_{\scriptscriptstyle 
BRS},f]\\
\label{eq:trentacinque}
\{f,g\}_{\scriptscriptstyle PB}=df[(dg)^{\sharp}] & \hat{\longrightarrow} & 
[[[f,Q_{\scriptscriptstyle BRS}],{\o K}],
[[[g,Q_{\scriptscriptstyle BRS}],{\o K}],K]]
\end{eqnarray} \vspace{0 mm} 
 \noindent The next thing to do is to translate the concept of Lie derivative 
which  is defined as: 
\begin{equation}
\label{eq:trentasei} 
{\cal L}_{\scriptscriptstyle V}=d\iota_{\scriptscriptstyle V}
+\iota_{\scriptscriptstyle V}d
\end{equation}
It is easy to prove that:
\begin{equation}
\label{eq:trentasette}
{\cal L}_{\scriptscriptstyle V}F^{(p)} \;\; \hat{\longrightarrow} \;\; 
i[{\widetilde {\cal H}}_{\scriptscriptstyle V},{\widehat F}
^{(p)}]
\end{equation} \vspace{0 mm} 

\noindent where ${\widetilde {\cal H}}_{\scriptscriptstyle 
V}=\lambda_aV^{a}+i\o{c}_a
\partial_bV^{a}c^{b}$. Note that,
for $V^{a}=\omega^{ab}\partial_bH$,  the ${\widetilde {\cal H}}_{\scriptscriptstyle 
V}$ becomes the ${\widetilde 
{\cal H}}$
of (\ref{eq:dodici}). This confirms that the full ${\widetilde {\cal H}}$ of eq.
(\ref{eq:dodici}) is the Lie derivative of the Hamiltonian flow .
One last point to notice is that  any $\HT$ associated to an Hamiltonian flow
can be written as:

\be
\label{eq:trentotto}
\HT=-i[\Qb[\QBb,H]]
\ee
\noindent
where $H$ is the 0-form out of which we build the Hamiltonian vector field
(via eq.(\ref{eq:trentaquattro})) which enters $\HT$. The structure 
of eq.({\ref{eq:trentotto}), that is
a double commutator with both BRS and antiBRS charges, is really what
embodies both the Lie-derivative structure and the Hamiltonian vector
field structure. A Lie derivative of a {\it vector} field (and not of
an Hamiltonian one) would have been expressed  only as a single commutator
(with respect to the BRS charge) and not as a double commutator. The first
commutator with the $\QBb$ in (\ref{eq:trentotto})  embodies the Hamiltonian
vector structure given by eq.(\ref{eq:trentaquattro}).

There are many  other structures in symplectic differential
geometry which can be translated in our formalism and the interested
reader can look them up in\scite{Geom}.

\noindent Besides the five conserved charges listed in eqs.(20)-(24)
there are two more \scite{Ergo}:

\begin{equation}
\label{eq:trentanove}
\NH=c^{a}\partial_{a}H~~~~~~;~~~~~~\NHB={\o c}_{a}\omega^{ab}\partial_{b}H
\end{equation}

\noindent Combining them with the $\Qb$ and $\QBb$ of (20)-(21) we get 
the following
two extra conserved charges:

\begin{equation}
\label{eq:quaranta}
\QH\equiv\Qb-\beta\NH~~~~~~~;~~~~~~~~\QBH\equiv\QBb+\beta{\overline\NH}
\end{equation}

\noindent where $\beta$ is a dimensional parameter. These two new charges
are true supersymmetry charges, in fact we have:
\begin{equation}
\label{eq:quarantuno}
[\QH,\QBH]=2i\beta\HT.
\end{equation}
\noindent We have not studied geometrically
 these charges in as many details as we did for the Isp(2) charges
 of eqs.(\ref{eq:venti})-(\ref{eq:ventiquattro}). That is what
 we plan to do in the next two sections.

\section{Gauging the Global Susy Invariance.}
\indent
The direction we take to study the geometrical structures behind the
supersymmetric charges above is to build a lagrangian where these symmetries
are local. The standard procedure we use is known in the literature\scite{Brin} 
as the Noether method.
It basically consists in finding the exact form of the extra terms
generated by {\it local} variations of the original lagrangian 
which had only the  global invariances. These extra terms, by Noether's
theorem, are basically  the derivatives of the infinitesimal parameters
multiplied by the generators. The trick then is to add to the original
lagrangian a piece made of an {\it auxiliary field} multiplied by
the generator. We can then impose that this auxiliary field  transform
in such a manner as to cancel the extra variations of the lagrangian
mentioned above.

As the susy charges are built out of the $\Qb, \QBb, \NH, \NHB$
let us build the {\it local} variations generated by each of these charges 
on the variables \quattrova. If we indicate with $X$ one of those four 
operators and with $(\cdot)$ any of the variables \quattrova, 
then by a local variation, 
$\delta^{\s loc}_{\scriptscriptstyle X}$, we indicate
the operation: $\delta^{\s loc}_{\scriptscriptstyle X}\equiv [\varepsilon(t)X,(\cdot)]$
where now the Grassmannian parameter $\varepsilon$ is dependent on~$t$.
These four variations are indicated below:

\be
\label{eq:quarantadue}
\begin{array}{ll}
	\delta^{\s loc}_{\scriptscriptstyle Q}\equiv
	\left\{
	\begin{array}{l}
	\delta\varphi^{a} = \epsilon(t)c^{a}\\
	\delta c^{a} = 0\\
	\delta \o{c}_{a} = i\epsilon(t)\l_{a}\\
	\delta \l_{a} = 0\\
	\end{array}
	\right. 
	&
	{\delta}^{\s loc}_{\o{\scriptscriptstyle Q}}\equiv
	\left\{
	\begin{array}{l}
	\delta\varphi^{a} = -{\o\epsilon}(t)\w^{ab}\o{c}_{b}\\
	\delta c^{a} = i{\o\epsilon}(t)\w^{ab}\l_{b}\\
	\delta \bc_{a} = 0\\
	\delta\l_{a} = 0
	\end{array}
	\right. 

\end{array}
\ee

\vskip 1cm

\be
\label{eq:quarantatre}
\begin{array}{ll}
	\delta^{\s loc}_{\scriptscriptstyle N}\equiv
	\left\{
	\begin{array}{l}
	\delta\varphi^{a} = 0\\
	\delta c^{a} = 0\\
	\delta \bc_{a} = \epsilon(t)\partial_{a}H\\
	\delta \l_{a} = i\epsilon(t)c^{b}\partial_{b}\partial_{a}H\\
	\end{array}
	\right. 
	&
	{\delta}^{\s loc}_{\o{\scriptscriptstyle N}}\equiv
	\left\{
	\begin{array}{l}
	\delta\varphi^{a} = 0\\
	\delta c^{a} = {\o\epsilon}(t)\w^{ab}\partial_{b}H\\
	\delta \bc_{a} = 0\\
	\delta\l_{a} = i{\o\epsilon}(t)\bc_{d}\w^{db}\partial_{b}\partial_{a}H.
	\end{array}
	\right. 

\end{array}
\ee

\noindent We could have used four different parameters for the four
different charges (as we will do later on)
 but here we limit ourselves just to two:
$\e(t)$ and ${\o\e}(t)$.
The local susy variations associated to the two susy charges of eq.
(\ref{eq:quaranta})
are :

\be
\label{eq:quarantaquattro}
	\left\{
	\begin{array}{l}
	\delta^{\s loc}_{\s{\QH}}=\delta^{\s loc}_{\s{Q}}-\beta\delta^{\s loc}_{\s{N}} \\
	{\delta}^{\s loc}_{\s\QBH}={\delta}^{\s loc}_{\o{\s Q}}+
	\beta{\delta}^{\s loc}_{\o{\s N}}. \\
	\end{array}
	\right.
\ee

\noindent It is now straightforward to check that the local susy variations 
of the lagrangian $\LT$~in (\ref{eq:sette}) gives the following results:

\be
\label{eq:quarantacinque}
	\delta^{\s loc}_{\s{\QH}}\widetilde{\mathcal L}=-i\dot{\e}\QH + (t.d.)
\ee

\noindent and

\be
\label{eq:quarantasei}
	{\delta}^{\s loc}_{\s{\QBH}}\widetilde{\mathcal L}=-i\dot{\o\e}\QBH + (t.d.).
\ee

\noindent With $(t.d.)$ we indicate total derivative terms. They turn
into surface terms in the action and they disappear if we
require that  $\epsilon(t)$  and ${\o\e}(t)$  be zero 
at the end points of integrations as we will do from now
on. To do things in a cleaner manner we should have actually
checked the invariance using the integrated charge as explained in
appendix~A. Anyhow from eq.(\ref{eq:quarantacinque}) and
(\ref{eq:quarantasei}) we see that the lagrangian does not change by a total
derivative so the two local susy transformations are not symmetries and we have to modify the 
lagrangian to find another one which is invariant. If we find it, then it must also
be invariant\scite{Brin} under the composition of  two local 
susy transformations which  we can prove
(see appendix B)  to be the sum of a local supersymmetry transformation 
plus a {\it local} time-translation generated by $\HT$. This last one
is not a symmetry
of $\LT$ and the lagrangian changes by a term proportional to $\HT$
multiplied by the time-derivative of the symmetry parameter, exactly as the
Noether theorem requires. The trick\scite{Brin} to get the
invariance is to add to $\LT$
some auxiliary fields multiplied by the charges under which $\LT$ is not
invariant. In our case the complete lagrangian turns out to be :

\be
\label{eq:quarantasette}
	\widetilde{\mathcal L}_{\s susy}\equiv\widetilde{\mathcal L}+
	\o{\psi}\QH+\psi\QBH+g\widetilde{\mathcal H},	
\ee

\noindent where $g(t),\psi(t),{\o\psi}(t)$ are three new fields 
(the last two of Grassmannian nature)  whose variations
under the local susy will be determined by the requirement that $\LT_{susy}$
be invariant under the local susy variations of eq.(\ref{eq:quarantaquattro}).
In detail we get:

\be
\label{eq:quarantotto}
	\delta_{\s{\QH}}\widetilde{\mathcal L}_{susy}=-i\dot{\e}\QH+(\delta_{\s{\QH}}g)\widetilde{\mathcal
	H} + (\delta_{\s{\QH}}\o{\psi})\QH + (\delta_{\s{\QH}}\psi)\QBH +
	\psi(2i\e\beta\widetilde{\mathcal H})
\ee

\noindent and we see that the following transformations of the 
variables $g,\psi,{\o\psi}$ make $\LT_{susy}$ invariant 
under the local transformation associated to $\QH$: 

\be
\label{eq:quarantanove}
	\left\{
	\begin{array}{l}
	\delta_{\s{\QH}}\o{\psi}=i\dot{\e} \\
	\delta_{\s{\QH}}\psi=0 \\
	\delta_{\s{\QH}}g=+2i\e\beta\psi.
	\end{array}
	\right.
\ee

\noindent For the variation under $\QBH$ we  get:

\be
\label{eq:cinquanta}
	\delta_{\s{\QBH}}\widetilde{\mathcal L}_{susy}=-i\dot{\o\e}\QBH+({\delta}_{\s
	{\o Q}_{H}}g)\widetilde{\mathcal H} + ({\delta}_{\s{\o Q}_{H}}{\o\psi})\QH +
	 ({\delta}_{\s{\o Q}_{H}}\psi)\QBH +
	{\o\psi}(2i{\o\e}\beta\widetilde{\mathcal H})
\ee

\noindent and we see that the following transformations of the 
variables $g,\psi,{\o\psi}$  make $\LT_{susy}$ invariant under the local
transformation associated to $\QBH$:

\be
\label{eq:cinquantuno}
	\left\{
	\begin{array}{l}
	{\delta}_{\s{\QBH}}\o{\psi}= 0\\
	{\delta}_{\s{\QBH}}\psi=i\dot{{\o\e}} \\
	{\delta}_{\s{\QBH}}g=+2i{\o\e}\beta\o{\psi}.
	\end{array}
	\right.
\ee

\noindent 
Last we should check how $\LT$ changes under a local time-reparametrization.
We have to do that because this reparametrization appears in the composition
of two local susy transformations (Appendix B). The action of the local time reparametrization
on the variables \quattrova is listed in formula (B.9) of appendix B.
Under those trasformations we can easily prove that 

\be
\label{eq:cinquantadue}
\delta \LT =-i\dot{\eta}\HT
\ee

\noindent where $\eta (t)$ is the time-dependent parameter of the transformation. 
Let us now use this result in the variation of $\LT_{susy}$ under
time reparametrization:

\be
\label{eq:cinquantatre}
	\delta\LT_{susy} = -i\dot{\eta}\HT  +\delta\o{\psi}\QH +\delta\psi\QBH
	+\delta g\HT.
\ee

\noindent We immediately notice that $\LTloc$ is invariant
under the local-time reparametrization if we transform the variables
 $(\psi,\o\psi,g)$ as follows 

\be
\label{eq:cinquantaquattro}
	\left\{
	\begin{array}{l}
	\delta\psi = \delta\o{\psi} = 0\\
	\delta g=i\dot{\eta}.
	\end{array}
	\right.
\ee

\noindent 
So, putting together all the  three local symmetries, (\ref{eq:quarantanove}),
(\ref{eq:cinquantuno}) and (\ref{eq:cinquantaquattro}), we can say that
$\LTloc$ (if we choose $\beta=1$ in eqs. (44)-(51)) has the following local invariance:

\be
\label{eq:cinquantacinque}
	\left\{
	\begin{array}{l}
	\delta\psi=i\dot{{\o\e}} \\
	\delta\o{\psi}= i\dot{\e} \\
	\delta g= i\dot{\eta}+2i({\o\e}\o{\psi}+\e\psi).
	\end{array}
	\right.
\ee

\noindent
It is not the first time that one-dimensional systems with local-susy 
have been built. The first work was the classic one of Brink et al.\scite{Brin}.
Later on people\scite{Alv} have played with the supersymmetric
Quantum Mechanical model (SUSY-QM) of Witten\scite{Witte} turning 
its global susy  into a local invariance. 
Regarding this model, the author of ref.\scite{Alv} 
pretended to show that the locally supersymmetric quantum mechanics 
was equivalent  to the  standard SUSY QM with only global
invariance. The proof was based on the fact that, via the
analog of the transformations ({\ref{eq:cinquantacinque}), it is possible
to bring the variables ($\psi,{\o \psi}, g)$ to zero and so, looking 
at eq.(\ref{eq:quarantasette}), this would imply that we can turn  
$\LT_{susy}$  into $\LT$.
This kind of reasoning is {\it misleading}. In fact, while it is easy to check
(see \break Appendix D) that it is possible to bring the $(\psi, \o\psi, g)$ to zero
via the transformations (\ref{eq:cinquantacinque}), it should
be remembered that the starting point was a gauge theory, $\LT_{susy}$, 
and the value zero for
the variables $(\psi, \o\psi, g)$ is equivalent to a particular choice
of gauge-fixing. Anyhow, the {\it physical} theory has to be gauge-fixing 
independent and this is achieved\scite{Teitel} by restricting the physical 
states via the
BRS charge associated to the local symmetries. So in the gauge-fixing where
the $(\psi, \o\psi, g)$ are zero the locally-supersymmetric QM\scite{Alv}
has the same action as  the globally supersymmetric theory\scite{Witte} but
we have to restrict the states to the physical ones which are basically
those annihilated by the symmetry charges. So the two systems,
the one with global susy and the one with local susy, are not equivalent
even if they seem to be so in a particular gauge-fixing. Their Hilbert
spaces are different even if the dynamics, in a particular gauge-fixing,
is the same.  Moreover, even at the
level of counting of degrees of freedom we shall show that, while
$\LT$ has $8n$ independent variables, $\LTloc$ has $8n-6$. To do this analysis
we should go through the business of studying the constraints associated
to the local symmetries of $\LTloc$ as we are going to do in what follows.

The standard procedure is the one of Dirac~\scite{Vinc} which we will
follow here in detail. Looking at $\LTloc$ we see that the primary
constraints are:
\be
\label{eq:cinquantasei}
	\left\{
	\begin{array}{l}
	\pipsi = 0 \\
	\pipsibar = 0 \\
	\pigi = 0
	\end{array}
	\right.
\ee

\noindent
where $\pipsi$, $\pipsibar$ and $\pigi$ are the momenta associated
to $\psi$, $\o\psi$ and $g$. The {\it canonical} Hamiltonian\scite{Vinc} 
is then the following:

\be
\label{eq:cinquantasette}
	\HT_{can.}=\HTloc=\HT-\psi\QBH-\bpsi\QH-g\HT
\ee

\noindent
while the {\it primary}\scite{Vinc}
(or total)  Hamiltonian is:

\be
\label{eq:cinquantotto}
	\HT_{\s
	P}=(1-g)\HT-\psi\QBH-\bpsi\QH+u_{1}\pipsi+u_{2}\pipsibar+u_{3}\pigi
\ee

\noindent
and it is obtained by adding to $\HT_{can.}$ the primary
constraints (\ref{eq:cinquantasei}) via the Lagrange multipliers 
$u_{1},u_{2},u_{3}$. Next we have to impose that the primary constraints 
do not change under the time evolution, i.e:

\be
\label{eq:cinquantanove}
	\begin{array}{lrc}
	[\pipsi,\HT_{\s P}] = 0, \;\; & [\pipsibar,\HT_{\s P}] = 0, \;\; &
	[\pigi,\HT_{\s P}] =0.
	\end{array}	
\ee

\noindent
We have used commutators here but it would have been more
correct to use the Extended-Poisson-Brackets. We did that just because
the two structures are isomorphic as explained in eq.(\ref{eq:sedici}).
In particular the (graded) commutators we need in (\ref{eq:cinquantanove})
are

\be
\label{eq:sessanta}
	\begin{array}{lrc}
	[\pipsi,\psi] = 1, \;\; & [\pipsibar,\bpsi] = 1, \;\; & [\pigi,g] =-i.
	\end{array}	
\ee

\noindent
Using them we get from (\ref{eq:cinquantanove}) the following set of
{\it secondary}\scite{Vinc} constraints:
 
\be
\label{eq:sessantuno}
	\left\{
	\begin{array}{l}
	\QBH = 0 \\
	\QH = 0 \\
	\HT = 0.
	\end{array}
	\right.
\ee

\noindent
At this point the careful reader could ask  which are
the operators  
generating the full set of transformations (\ref{eq:cinquantacinque}),
expecially the last one. It does not seem
that they are generated by the operators of eqs. (\ref{eq:sessantuno})
and ({\ref{eq:cinquantasei}). Actually the answer to this question
is rather subtle and tricky \scite{Teitel} and is given in full details
in appendix C.

Having clarified this point, we can go on with our procedure.
We have now to require that also the secondary constraints
(\ref{eq:sessantuno}) do not change
under time evolution using as operator of evolution always the primary
Hamiltonian $\HT_{\s P}$ as explained in ref.\scite{Vinc}.
It is easy to realize that in our case we do not generate 
further constraints with this procedure and 
that, at the same time, we do not determine 
the Lagrange multipliers. The fact that the Lagrange multipliers are all 
left undetermined is a signal that the constraints are first 
class~\scite{Vinc}
as it is easy to check by doing the commutators among all the
six constraints (\ref{eq:sessantuno}) and (\ref{eq:cinquantasei}).
Being them first class, one has to introduce six gauge-fixings
which will be used to determine the Lagrange multipliers\scite{Vinc}.

The gauge-fixings, let us call them  $\chi_{i}$, must 
have a non-zero commutator with the
associated gauge generator. For the three constraints of
eq.(\ref{eq:cinquantasei}) three suitable gauge-fixings can be:

\be
\label{eq:sessantadue}
	\begin{array}{lrc}
	\psi-\psi_{\s 0} = 0, \;\; & \bpsi-\bpsi_{\s 0} = 0, \;\; & g-g_{\s 0} =
	0,
	\end{array}	
\ee

\noindent

\noindent
where $\psi_{\scriptscriptstyle 0}$, ${\o\psi}_{\scriptscriptstyle 0}$ 
and  $g_{\scriptscriptstyle 0}$ are three fixed functions. It is easy to check that 
each of them does not commute with
its associated generator. The gauge-fixing $\psi_{\scriptscriptstyle 0}=
{\o\psi}_{\scriptscriptstyle 0}=g_{\scriptscriptstyle 0}=0$ 
is among the admissible ones, in the sense that there is a gauge 
transformation which
brings any configuration into this one as shown in appendix D.
 In this gauge fixing
we get that the $\gaug$ variables disappear from the $\LTloc$
and so $\LTloc$  apparently is reduced to $\LT$. Of course, as we said earlier,
this should not mislead us to think that the physics
of $\LTloc$ is the same as the one of $\LT$. In fact 
at the Hamiltonian level, even if $\HTloc$ is reduced to $\HT$
by the gauge-fixing, the Poisson brackets for the two systems are different. 
For the one with local symmetries the Poisson brackets are the Dirac ones
which, given two observables $O_{1}$ and $O_{2}$, are built as:

\be
\label{eq:sessantatre}
	\{O_{1},O_{2}\}_{\s DB} = \{O_{1},O_{2}\} -
	\{O_{1},G_{i}\}(C^{-1})^{ij}\{G_{j},O_{2}\},
\ee
\noindent
We have indicated with $G_{i}$ any of the six first class
constraints of eq.(\ref{eq:sessantuno})(\ref{eq:cinquantasei}),
and the matrix $C_{ij}$ has its elements built as $\{G_{i},\chi_{j}\}$
where $\chi_{i}$ are the six gauge-fixings  associated to
the constraints $G_{i}$. The brackets entering the expressions
on the RHS of (\ref{eq:sessantatre})
are the standard Extended Poisson Brackets of eq.({\ref{eq:tredici}).
If the dynamics is  the one of a system with global susy only,
that is one whose Hamiltonian is really $\HT$ from the beginning,
then the Poisson brackets among the  same two observables 
$O_{1}$ and $O_{2}$ would be $\{O_{1},O_{2}\}$ and it is clear that

\begin{equation}
\label{eq:sessantaquattro}
	\{O_{1},O_{2}\} \neq \{O_{1},O_{2}\}_{\s DB}.
\end{equation}

\noindent
This explains why, even if the Hamiltonians of the two systems
(the one with local symmetries and the one with global ones) are the same
(in some gauge-fixings), the two dynamics are anyhow different because they
are ruled by different Poisson brackets.

Also the counting of the degrees of freedom indicates that the
systems have different numbers of degrees of freedom. The one with global
symmetries, and lagrangian $\LT$, has just the variables \quattrova which 
are $8n$.
The one with local symmetries, $\LTloc$, has the variables \quattrova
which are $8n$, plus the three gauge variables $\gaug$ and 
the relative momenta for a total of
6, minus the 6 constraints of eq.(\ref{eq:sessantuno})(\ref{eq:cinquantasei}),
minus the 6 gauge fixings $\chi_{i}$, for a total of $8n-6$ variables.
This is the correct counting of variables as explained in 
ref.\scite{Teitel}. So even from this we realize that the
two systems are different. 

As we said at the beginning, our path-integral is actually the counterpart
of the operatorial version of CM proposed by 
Koopman and von Neumann\scite{Koop}  and if we adopt this operatorial version
we should use the commutators derived in (\ref{eq:quindici}).
This operatorial version of course could be adopted also for the dynamics
with local symmetries associated to~$\HTloc$ . In the operatorial
formulation we have a Hilbert space but we know that, for a system
with local symmetries,
the Hilbert space is restricted to the {\it physical
states}. The selection of these states is done by a BRS-BFV charge\scite{Teitel}
associated to the gauge-symmetries of the system. This BRS-BFV charge 
of course  has nothing
to do with the $\Qb$ of eq.(\ref{eq:venti}). The construction of the 
BRS-BFV
charge for our $\LTloc$ goes as follows\scite{Teitel}. First we should introduce
a pair of {\t gauge} ghost-antighosts for each gauge generator. 
As our gauge generators are
 
\be
\label{eq:sessantacinque}
	G_{i}=(\pipsibar, \pipsi, \pigi, \QH, \QBH, \HT)
\ee

\noindent
the ghost-antighosts are twelve and can be indicated as:

\be
\label{eq:sessantasei}
	\begin{array}{l}
	\eta^{i} = (\eta_{\bpsi}, \eta_{\psi}, \eta_{g}, \eta_{\s H},
	\o{\eta}_{\s H}, \widetilde{\eta}_{\s H})
	\\
	{\mathcal P}_{i} = ({\mathcal P}_{\bpsi}, {\mathcal P}_{\psi}, {\mathcal
	P}_{g}, {\mathcal P}_{\s H},
	\o{\mathcal P}_{\s H}, \widetilde{\mathcal P}_{\s H}).
	\end{array}
\ee

\

\noindent
The general BRS-BFV charge\scite{Teitel} is then\footnote{The graded commutators
among the ghosts of (\ref{eq:sessantasei}) are $[\eta^{i},{\mathcal
P}_{j}]=\delta^{i}_{j}$}: 
 
\be
\label{eq:sessantasette}
	\Omega_{\s BRS} = \eta^{i}G_{i} -
	\frac{1}{2}(-)^{\e_{i}}\eta^{i}\eta^{j}C_{ji}^{k}{\mathcal P}_{k},
\ee

\noindent
where  $\varepsilon_{\s i}$ is the Grassmannian grading of the
constraints $G_{i}$ and $C^{k}_{ij}$ are the structure constants
of the algebra of our constraints. In our case this algebra is:

\be
\label{eq:sessantotto}
	\begin{array}{l}
	\,[\QH,\QBH] = 2i\HT \\
	\,[\QH,\QH] = [\QBH,\QBH] = [\QH,\HT] = [\QBH,\HT] = 0 
	\end{array}
\ee
\noindent
where we have put $\beta=1$ with respect to eq.(\ref{eq:quarantuno}).

\noindent
It is now easy to work out the BRS-BFV charge for our  local
susy invariance:

\be
\label{eq:sessantanove}
	\Omega_{\s BRS}^{(susy)}=\eta_{\bpsi}\pipsibar+\eta_{\psi}\pipsi+\eta_{g}\pigi+\eta_{\s
	H}\QH+
	\o{\eta}_{\s H}\QBH+\widetilde{\eta}_{\s H}\HT-2i\eta_{\s
	H}\o{\eta}_{\s H}
	\widetilde{\mathcal P}_{\s H}	
\ee

\noindent
Note that it contains terms with three ghosts and so it is
hard to see how it works on the states. These terms with three
ghosts are there because the generators are not in involution.
As it is explained in ref.\scite{Teitel} in case the constraints
$G_{i}$  are not in involution, one can build some new ones
$F_{i}=a^{j}_{i}G_{j}$ which are in involution. In our case the $F_{i}$
generators, replacing the $\QH$ and $\QBH$,
can be easily worked out and they  are:

\be
\label{eq:settanta}
	\left\{
	\begin{array}{l}
	Q_{\s A}\equiv(\QH + \psi\HT) \\
	Q_{\s B}\equiv(\QBH - 2i\pipsi) 
	\end{array}
	\right.
\ee

\noindent
while the other $F_{i}$ are the same as the $G_{j}$. The associated 
BRS-BFV charge is then

\be
\label{eq:settantuno}
	\Omega^{\s (F)}_{\s
	BRS}=\eta_{\bpsi}\pipsibar+\eta_{\psi}\pipsi+\eta_{g}\pigi+\eta_{\s
	A}Q_{\s A}+
	\eta_{\s B}Q_{\s B}+\widetilde{\eta}_{\s H}\HT.		
\ee
\noindent
We have called $\eta_{\s A}, \eta_{\s B}$ the BFV ghosts
associated to $Q_{\s A}, Q_{\s B}$. Note that this $\Omega^{\s (F)}_{\s BRS}$ does not
contain terms with three ghosts. The {\it physical states} are then defined
as

\begin{equation}
\label{eq:settantadue}
	\Omega^{\s (F)}_{\s BRS}\mid\mbox{phys}\rangle = 0
\end{equation}

\noindent
and, by following ref.\scite{Teitel}, we can easily show that 
(\ref{eq:settantadue}) is equivalent to the following six constraints:

\be
\label{eq:settantatre}
	\begin{array}{ll}
	(1)\left\{
	\begin{array}{l}
	\pipsibar\mid\mbox{phys}\rangle = 0 \\
	\pipsi\mid\mbox{phys}\rangle  = 0 \\
	\pigi\mid\mbox{phys}\rangle  = 0
	\end{array}
	\right. 
	\;\;&\;\;
	(2)\left\{
	\begin{array}{l}
	Q_{\s A}\mid\mbox{phys}\rangle = 0 \\
	Q_{\s B}\mid\mbox{phys}\rangle = 0 \\
	\HT\mid\mbox{phys}\rangle = 0.
	\end{array}
	\right.
	\end{array}
\ee

\noindent
The set (1) above means that the physical states must be independent
of  $\gaug$ that means independent of any choice of gauge fixing.
The set (2) instead (combined with some of the  conditions from the 
set (1)) is equivalent to the following conditions:

\begin{equation}
\label{eq:settantaquattro}
	\begin{array}{l}
	\QH\mid\mbox{phys}\rangle = 0 \\
	\QBH\mid\mbox{phys}\rangle = 0 \\
	\HT\mid\mbox{phys}\rangle = 0.
	\end{array}
\end{equation}

\noindent
We can summarize it
by saying that, even if in some gauge-fixing
the $\HTloc$ is the same as $\HT$, the dynamics of the first
is restricted to a subset (given by eq.(\ref{eq:settantaquattro}))
of the full Hilbert space while the dynamics of $\HT$ is not restricted.
This is what is not spelled out correctly in ref.\scite{Alv}.
The author may have been brought to the wrong conclusions
not only because he  did not consider the correct Hilbert space
but also by the following fact that we like to draw to the attention
of the reader. The three quantities $\QH,\QBH,\HT$
entering the constraints (\ref{eq:sessantuno})
are actually constants of motion in the space \quattrova. So fixing
them, as the constraints do, is basically fixing a set of initial 
conditions. The hypersurface defined by (\ref{eq:sessantuno})
is the subspace of \quattrova where the motion takes place and 
it takes place  with the same dynamics as in \quattrova. 
If the constraint surfaces were not made by constants of 
motion, then the dynamics would have to be  modified to force the
particle to move on them, but this is not the case here.

Before concluding let us notice that 
the analysis we have made  seems to  tell us that, by forcing the particle 
to move on some hypersurfaces fixed by particular values of the constants
of motion, we  generate a dynamics with  local-invariances. We will
explore this issue further in later sections of the paper.

\section{New Local Susy Invariance and Equivariance }

\indent
We have seen in  section 2 that, besides the susy, there are other
global invariances of $\HT$. We are tempted to gauge all of them and see
what comes out. In this paper we will limit ourselves to study what
happens  when we gauge separately the $\Qb$, $\QBb$, $N_{\s H}$, 
${\o N}_{\s H}$ of eqs. (\ref{eq:venti}), (\ref{eq:ventuno}) and
(\ref{eq:trentanove}). The local
variation of $\LT$ under these four combined gauge transformations is 

\bea
\label{eq:settantacinque}
\delta^{loc.}\LT & = & [\e \Qb+{\o\e}\QBb+\eta N_{\s H}+{\o\eta}{\o N}_{\s H},\LT]
\nonumber\\
& = & -i{\dot\e}\Qb-i{\dot{\o\e}}\QBb-i{\dot\eta}N_{\s H}-i{\dot{\o\eta}}{\o
N}_{\s H}
\eea

\noindent
where $(\e(t),{\o\e}(t),\eta(t),{\o\eta}(t))$ are  four different
Grassmannian gauge
parameters. From equation (\ref{eq:settantacinque}) one could be tempted to propose
the following as {\it extended} lagrangian invariant under the local symmetries 
above:

\be
\label{eq:settantasei}
\LT_{ext.}\equiv\LT+\alpha(t)\Qb+{\o\alpha}(t)\QBb+\beta(t)N_{\s H}+{\o\beta}(t)
{\o N}_{\s H}
\ee
\noindent
where $(\alpha(t),{\o\alpha}(t),\beta(t),{\o\beta}(t))$ are four Grassmannian
gauge-fields which we could transform in a proper way in order to 
make $\LT_{ext.}$ invariant.
This is actually impossible whatever transformation we envision
for the gauge-fields. In fact, as we did in the case of the susy of 
the previous section, we should consider what happens when we compose two 
of the local symmetries of eq.(\ref{eq:settantacinque}). This information
is given by the following commutators\scite{Ennio}:

\be
\label{eq:settantasette}
[\Qb,{\o N}_{\s H}]=i\HT~~~~~~~~~;~~~~~~~~~~[\QBb,N_{\s H}]=-i\HT
\ee

\noindent
This basically tells us that we should add to the lagrangian $\LT_{ext.}$ of
eq.(\ref{eq:settantasei}) an extra gauge field $g(t)$ and an extra gauge
generator\footnote{We will indicate with greek letters the gauge fields
associated to Grassmannian generators and with latin letters the one
associated to bosonic generators.} $\HT$:

\be
\label{eq:settantotto}
\LT_{ext.}=\LT+\alpha(t)\Qb+{\o\alpha}(t)\QBb+\beta(t)N_{\s H}+{\o\beta}(t)
{\o N}_{\s H}+g(t)\HT
\ee

\noindent
Doing now an extended gauge transformation like in (\ref{eq:settantacinque})
we get:

\bea
\label{eq:settantanove}
\delta_{loc.}\LT_{ext.}= & - & i{\dot\e}\Qb-i{\dot{\o\e}}\QBb 
-i{\dot\eta} N_{\s H}- i{\dot{\o\eta}}{\o N}_{\s H}+(\delta\alpha)\Qb\nonumber\\
& + & i\alpha{\o\eta}\HT+(\delta{\o\alpha})\QBb-i{\o\alpha}\eta\HT
+(\delta\beta) N_{\s H}\nonumber\\
& - & i\beta{\o\e}\HT+(\delta{\o\beta})~{\o N}_{\s H}+i{\o\beta}\e\HT +
(\delta g)\HT 
\eea

\noindent
and from this it is easy to see that $\LT_{ext.}$ is invariant 
if the gauge-fields $(\alpha,{\o\alpha}, \beta, {\o\beta}, g)$ 
are transformed  as follows:

\be
\label{eq:ottanta}
\left\{
\begin{array}{l}
\delta\alpha=i{\dot\e}\\
\delta{\o\alpha}= i{\dot{\o\e}}\\
\delta\beta=i{\dot\eta}\\
{\delta}{\o\beta}=i{\dot{\o\eta}}\\
\delta g = i{\o\alpha}\eta-i\alpha{\o\eta}+i\beta{\o\e}-i{\o\beta}\e
\end{array}
\right.
\ee
\noindent
From these transformations we notice that, for some choice of the gauge-fields
and of the gauge-transformations, we do not need to have the $g\HT$ in 
the $\LT_{ext.}$
The first choice is $\alpha={\o\alpha}=\e={\o\e}=0$
which, from eq.(\ref{eq:ottanta}), implies that we can choose $g(t)=0$. The
$\LT_{ext.}$ is then

\be
\label{eq:ottantuno}
\LT_{\s N}\equiv\LT+\beta(t)N_{\s H}+{\o\beta}(t){\o N}_{\s H}
\ee

\noindent
The second choice, which also implies that we can choose $g(t)=0$, is 
$\beta={\o\beta}=\eta={\o\eta}=0$
and this would lead to the following lagrangian

\be
\label{eq:ottantadue}
\LT_{\s \Qb}\equiv\LT+\alpha(t)\Qb+{\o\alpha}(t)\QBb
\ee

\noindent
We shall hang around  here for a moment  spending some time
on the lagrangian $\LT_{\s \Qb}$ above
because it allows us to do some  crucial observations on the counting 
of the degrees of freedom. As we said in the previous
section the lagrangian $\LT_{susy}$ of eq.(\ref{eq:quarantasette}) has 
fewer degrees of freedom than the standard one $\LT$ of eq.(\ref{eq:sette}), 
and the same happens with ~$\LT_{\s \Qb}$. In fact in $\LT_{\s \Qb}$~ we have
two primary constraints:

\be
\label{eq:ottantatre}
\Pi_{\s\alpha}=0~~~;~~~\Pi_{\s {\o\alpha}}=0
\ee

\noindent
which generate two secondary ones:

\be
\label{eq:ottantaquattro}
\Qb=0~~~;~~~\QBb=0
\ee

\noindent
All these four constraints are first class so we need four gauge-fixings.
The total counting\scite{Teitel} is then $8n$ (original variables) $+ 4$ (gauge variables and momenta)
$-4$ (constraints) $-4$ (gauge-fixings) for a total of $8n-4$ phase-space variables. 

\noindent
At this point one question that arises naturally is: ``{\it Is it possible to have a lagrangian
with the local invariances generated by $\Qb$ and $\QBb$, but with the same number
of degrees of freedom as }$\LT$?". The answer is yes. In fact let us  start from the following
lagrangian:

\be
\label{eq:ottantacinque}
\LT_{\s \Qb}^{\prime}\equiv\LT-{\dot\alpha}(t)\Qb-{\dot{\o\alpha}}(t)\QBb
\ee

\noindent
It is easy  to check that it is invariant under the following set of
local transformations generated by the $\Qb$ and $\QBb$:
\bea
\label{eq:ottantasei}
\delta(\cdot) & = & [\e(t)\Qb+{\o\e}(t){\QBb},(\cdot)]\nonumber\\
\delta\alpha & = &-i{\e}\nonumber\\
\delta{\o\alpha} & = &-i{\o\e}
\eea

\noindent
where we have indicated with $(\cdot)$ any of the variables \quattrova.
Anyhow at the same time it is easy to check that the lagrangian 
$\LT^{\prime}_{\s \Qb}$ of (\ref{eq:ottantacinque}) has only two primary constraints 
\bea
\label{eq:ottantasette}
\Pi_{\s\alpha} & = & {\Qb}\nonumber\\
\Pi_{\s{\o\alpha}} & = & {\QBb}
\eea

\noindent
and no secondary ones. The above two constraints are first class so we need
just two gauge-fixings and not four as before. The counting of degrees of
freedom now goes as follows: $8n$ (original variables)$+4$(gauge variables and
momenta)$-2$ (constraints)$- 2$ (gauge fixings) for a total of $8n$ variables. So we see that the system 
described by the lagrangian
$\LT^{\prime}_{\s\Qb}$ of eq.(\ref{eq:ottantacinque}) has the same number
of degrees of freedom as the original $\LT$. In appendix E we will show how
the constraints (87) act in the Hilbert space of the system. The fact that
$\LT^{\prime}_{\s\Qb}$ and $\LT$ are somehow equivalent could also
be understood by doing an integration by part of the terms 
of eq.(\ref{eq:ottantacinque})  containing
${\dot\alpha}$ and ${\dot{\o\alpha}}$. The integration by parts produces,
with respect to $\LT$, some terms which vanish  because of the conservation of
$\Qb$ and $\QBb$.

Even for the invariance under local susy of the previous section a mechanism
like the one above could be implemented. One  just needs to add to $\LT$ terms
like those in eq.(\ref{eq:quarantasette}) but with the gauge fields replaced
by their derivatives. This we feel may have been what has happened in 
ref.\scite{Zanel} where the authors have a system with local susy which 
has anyhow the same effective number of degrees of freedom as the model 
with global susy.

Going back  to the lagrangian (\ref{eq:ottantacinque}), we should mention
that it is not the first time  people thought of making the BRS-antiBRS
invariance local\scite{Baul}. We will not
expand this issue here because we want to stick to the susy symmetry.
We will come back in ref.\scite{Hilb} to the issue of gauging the BRS
symmetry and all the Isp(2) charges of
eqs.(\ref{eq:venti})-(\ref{eq:ventiquattro}).

Let us now return to eq.(\ref{eq:settantotto}). The two choices
which led to eq.({\ref{eq:ottantuno}) and ({\ref{eq:ottantadue})
are not the only ones consistent with the transformations (\ref{eq:ottanta}).
Another choice is

\be
\label{eq:equi-uno}
\begin{array}{ll}
	\left\{
	\begin{array}{l}
	\alpha=-{\o\beta}\\
	{\o\alpha}=\beta
	\end{array}
	\right. 
	&
	\left\{
	\begin{array}{l}
	\e=-{\o \eta}\\
	{\o\e}=\eta ~~.
	\end{array}
	\right. 

\end{array}
\ee
\noindent
With this choice the lagrangian that we get from (\ref{eq:settantotto})
is:

\be
\label{eq:equi-due}
\LT_{eq.}\equiv \LT +\alpha(t) Q_{\s (1)}+{\o\alpha}(t)Q_{\s (2)}+g(t)\HT
\ee

\noindent
The suffix $(eq.)$ on the lagrangian is for ``equivariant" and the 
reason will be clear later on. The $Q_{\s (1)}, Q_{\s (2)}$ on the RHS of eq.
$(\ref{eq:equi-due})$ are

\be
\label{eq:equi-tre}
	\left\{
	\begin{array}{l}
	Q_{\s (1)}\equiv\Qb-{\o N}_{\s H}\\
	Q_{\s (2)}\equiv\QBb+N_{\s H} 
	\end{array}
	\right.
\ee
\noindent
Note that these charges, with respect to the $\QH$ and $\QBH$ 
of eq.(\ref{eq:quaranta}), are somehow twisted in the sense that 
here we sum the $\Qb$ with ${\o N}$
and not with $N$ and viceversa for the $\QBb$.
It is easy to check that:

\be
\label{eq:equi-quattro}
Q^{2}_{\s (1)}=Q^{2}_{\s (2)}=-i\HT~~~;~~~[Q_{\s (1)},Q_{\s (2)}]=0
\ee

\noindent
So these two charges generate two supersymmetry transformations
which are anyhow different from those generated by the supersymmetry
generators of eq.(\ref{eq:quaranta}). The lagrangian of
eq.(\ref{eq:equi-due}) has two {\it local} susy invariances but different 
from the ones of $\LT_{susy}$ (\ref{eq:quarantasette}). 
In order to get the lagrangian 
(\ref{eq:quarantasette}) we should have made in eq.(\ref{eq:settantotto})
and ({\ref{eq:ottanta}) the following choice:

\be
\label{eq:equi-cinque}
\begin{array}{ll}
	\left\{
	\begin{array}{l}
	\alpha=-\beta={\o\psi}\\
	{\o\alpha}={\o\beta}=\psi
	\end{array}
	\right. 
	&
	\left\{
	\begin{array}{l}
	\e=-\eta\\
	{\o\e}={\o\eta}.
	\end{array}
	\right. 

\end{array}
\ee

\noindent
Going back to eq.(\ref{eq:equi-due}), let us now restrict the lagrangian
to the following one:

\be
\label{eq:equi-sei}
\LT_{eq.}=\LT +\alpha(t)Q_{\s (1)}+g(t)\HT
\ee
\noindent
which is locally invariant only under one susy and the symmetry
transformations are:

\be
\label{eq:equi-sette}
\left\{
\begin{array}{l}
\delta (\cdot)  =  [\e Q_{\s (1)}+\tau\HT, (\cdot)]\\
\delta\alpha  =  i{\dot\e}\\
\delta g  =  2i\alpha\e+i{\dot\tau}
\end{array}
\right.
\ee
\noindent
where $(\cdot)$ indicates any of the variables \quattrova
and $\e(t)$ and $\tau(t)$ are infinitesimal parameters.
Because of this gauge invariance, we have to handle the system
either via the Faddeev procedure\scite{Pavao} 
or the BFV method\scite{Teitel}. We will follow this last one.
The constraints (primary and secondary) derived from (\ref{eq:equi-sei}) are:

\be
\label{eq:equi-otto}
\left\{
\begin{array}{l}
\Pi_{\s \alpha}=0\\
\Pi_{\s g}=0\\
Q_{\s (1)}=0\\
\HT=0
\end{array}
\right.
\ee

\noindent
where $\Pi_{\s\alpha}$ and $\Pi_{\s g}$ are respectively the momenta
conjugate to the gauge fields $\alpha(t)$ and $ g(t)$. The BFV procedure,
as explained in the previous section, tells
us to add four new ghosts and their respective momenta to the system. We will
indicate them as follows:

\be
\label{eq:equi-nove}
\left\{
\begin{array}{l}
(C^{\s (1)}, C^{\s H}, {\o C}_{\s (1)}, {\o C}_{\s H})\\
({\o{\cal P}}_{\s (1)}, {\o{\cal P}}_{\s H},{\cal P}_{\s (1)},{\cal P}_{\s H})
\end{array}
\right.
\ee\noindent
We shall impose the following graded-commutators:

\be
\label{eq:equi-dieci}
        \left\{
	\begin{array}{l}
	\,[g,\Pi_{\s g}]=[C^{\s(1)},{\o\mathcal{P}}_{\s(1)}]=
	[{\o C}_{\s(1)},\mathcal{P}_{\s(1)}]=1\\
	\,[\alpha,\Pi_{\s\alpha}]=[C^{\s H},{\o{\mathcal P}}_{\s H}]=
	\,[{\o C}_{\s H},{\mathcal P}_{\s H}]=1
	\end{array}
	\right.
\ee
In the first line above the variables are all ``bosonic" while
in the second one are all Grassmannian.
Equipped with all these tools  we will now build the
BFV-BRS\scite{Teitel} charge associated to our constraints:

\be
\label{eq:equi-undici}
\Omega_{\s BRS}^{\s (eq.)}\equiv C^{\s (1)}Q_{\s (1)}+C^{\s H}\HT+{\mathcal P}_{\s (1)}
\Pi_{\s\alpha}+{\mathcal P}_{\s (H)}\Pi_{\s g}+i (C^{\s (1)})^{2}
{\o{\mathcal P}}_{\s H}
\ee

\noindent
It is easy to check that $(\Omega_{\s BRS}^{\s (eq.)})^{2}=0$ as a BRS charge
should be. The next step, analogous to what we did in section 3, is to
select as physical states those annihilated by the $\Omega_{\s BRS}^{\s (eq).}$
charge:

\begin{equation}
\label{eq:equi-dodici}
	\Omega^{\s (eq.)}_{\s BRS}\mid\mbox{phys}\rangle = 0
\end{equation}

\noindent
Because of the nilpotent character
of the $\Omega^{\s (eq.)}_{\s BRS}$, we should remember that two 
physical states 
are equivalent if they differ by a BRS variation:

\be
\label{eq:equi-tredici}
\mid\mbox{phys-2}\rangle=\mid\mbox{phys-1}\rangle+
\Omega^{\s (eq.)}_{\s BRS}\mid\mbox{$\chi$}\rangle 
\ee

\noindent
Performing the standard procedure\scite{Teitel} of abelianizing the constraints
(\ref{eq:equi-otto}) and building the analog of the $\Omega_{\s BRS}^{(F)}$ 
of eq.(\ref{eq:settantuno}), it is then easy to see that the physical state
condition $(\ref{eq:equi-dodici})$ is equivalent to the following four conditions:

\be
\label{eq:equi-quattordici}
\left\{
\begin{array}{l}
\HT\mid\mbox{phys}\rangle=0\\
Q_{\s (1)}\mid\mbox{phys}\rangle=0\\
\Pi_{\s \alpha}\mid\mbox{phys}\rangle=0\\
\Pi_{\s g}\mid\mbox{phys}\rangle=0
\end{array}
\right.
\ee

\noindent
Let us now pause for a moment and, for completeness,
let us  briefly review the concept
of equivariant cohomology (for references see\scite{Berline}).
Let us indicate with $\psi$ and $\chi$ two {\it inhomogeneus} forms on
a symplectic space and with $V$ a vector field on the same  space.
One says that the form $\psi$ is equivariantly closed but not exact
with respect to the vector field $V$ if the following  conditions
are satisfied:

\be
\label{eq:equi-quindici}
\left\{
\begin{array}{l}
{\cal L}_{\s V}\psi=0\\
{\cal L}_{\s V}\chi=0\\
(d-\iota_{\s V})\psi=0\\
\psi\neq(d-\iota_{\s V})\chi
\end{array}
\right.
\ee
\noindent
The forms $\psi$ and $\chi$  have to be inhomogeneous
because, while the exterior derivative~$d$ increases the degree
of the form of one unit, the contraction with the vector
field $\iota_{\s V}$ decreases it of one unit, so the third
and fourth relations in eq.(\ref{eq:equi-quindici}) would never have
a solution if $\psi$ and $\chi$ were homogeneous in the form degree.
From now on let us use the notation: $d_{eq.}\equiv (d-\iota_{\s V})$
and let us try to interpret the  relations contained in
eq.({\ref{eq:equi-quindici}).
Restricting the forms to satisfy the first two relations contained in
({\ref{eq:equi-quindici}) and noting that $d_{eq.}^{2}= -{\cal L}_{\s V}$,
we  have that on this restricted space $[d_{eq.}]^{2}=0$, and so 
$d_{eq.}$ acts
as an exterior derivative. If we  now consider the last two relations of 
(\ref{eq:equi-quindici}) it is then clear that they  define a cohomology 
problem for $d_{eq.}$.

There are other more abstract definitions of
equivariant  cohomology\scite{Cart} based on the {\it basic} cohomology of the
Weil algebra associated to a Lie-algebra, but we will not dwell on it here.
Equivariant cohomology is a concept that entered also the famous localization 
formula of Duistermaat and Heckman\scite{Duit} thanks to the work of Atiyah and Bott
and into Topological Field Theory thanks to the work of R. Stora
and collaborators\scite{Topol}.

Let us now go back to our lagrangian $\LT_{eq.}$ of eq. (\ref{eq:equi-sei})
whose physical state space is
restricted by the conditions (\ref{eq:equi-quattordici}) because
of the gauge invariance given in (\ref{eq:equi-sette}). It is easy to
realize that the first two physical state conditions of  
(\ref{eq:equi-quattordici}) are equivalent to the  first and third conditions
of eq. (\ref{eq:equi-quindici}) once the vector field $V$ is identified
with the Hamiltonian vector field $(dH)^{\sharp}$~\scite{Marsd}. In fact
let us remember the correspondence described 
in section 2 between standard operations in differential geometry
and in our formalism and in particular formula (\ref{eq:trentasette}).
This tells us that  the Lie-derivative ${\cal L}_{\s {dH}^{\sharp}}$ acts
on a form  as the commutator of $\HT$ with the same form written in terms of
$c^{a}$ variables. This commutator gives the same result as the action
of $\HT$  on functions of $\varphi^{a}$ 
and $c^{a}$ once $\HT$ is  written as a differential 
operator like in (\ref{eq:equi-quattordici}). So 
eq. (\ref{eq:trentasette}) proves that the first condition in both eqs.
(\ref{eq:equi-quindici}) and 
(\ref{eq:equi-quattordici}) is the same: 

\be
\label{eq:equi-sedici}
{\cal L}_{\s ({dH}^{\sharp})}\psi=0\longrightarrow\HT\mid\mbox{phys}\rangle=0
\ee

\noindent
Next let us look at the second condition in (\ref{eq:equi-quattordici})
and the third in (\ref{eq:equi-quindici}). 
From the form (\ref{eq:equi-tre}) of $Q_{\s (1)}$ we see that its first term,
the $\Qb$,  via the correspondence given by eq.(\ref{eq:ventinove}), 
corresponds to the exterior derivative $d$ which is exactly the first term 
contained in the third relation of eq.(\ref{eq:equi-quindici}). 
The second term in  $Q_{\s (1)}$ is the ${\o N}$ which is given in 
eq.(\ref{eq:trentanove}) and can be written as

\be
\label{eq:equi-diciassette}
{\o N}_{\s H}=[\QBb, H]
\ee

\noindent
From the relation (\ref{eq:trentaquattro}) we see that we can
interpret ${\o N}_{\s H}$ as the Hamiltonian vector field built out of
the function $H$. Its action as a differential
operator on forms is then given by eq.(\ref{eq:trenta}), that means it
acts  as the interior contraction, $\iota_{\s
({dH}^{\sharp})}$, of the
Hamiltonian vector field with forms. This basically proves
the correspondence between the second relation of (\ref{eq:equi-quattordici})
and the third of (\ref{eq:equi-quindici}):

\be
\label{eq:equi-diciotto}
(d-\iota_{\s ({dH}^{\sharp})})\psi=0\longrightarrow Q_{\s (1)}\mid\mbox{phys}\rangle=0
\ee

\noindent
Note that this correspondence would not hold if we had gauged the $\QH$, like we
did in section 3. In fact the $\QH$, being made of $\Qb$ and $N_{H}$ and not
${\o N}_{H}$, would not have had the meaning of equivariant exterior derivative.

Let us now conclude the proof that the conditions (101) are really equivalent to
the equivariant
cohomology problem given by eq.~(102). We have already explained in eqs.~(103)
and (105) the
correspondences:
	\[
	\begin{array}{c}
	{\cal L}_{\s V}\psi=0\longleftrightarrow\HT\mid\mbox{phys}\rangle=0 \\
	(d-\iota_{\s V})\psi=0\longleftrightarrow Q_{\s
	(1)}\mid\mbox{phys}\rangle=0. 
	\end{array}
	\]
\noindent We haven't discussed yet the 2nd and 4th equations in (102). These
conditions are equivalent to the following
statement:
	\be
	\{\psi=(d-\iota_{\s V})\chi ~~~\mbox{with}~~~{\cal L}_{\s
	V}\chi=0\}~~\Longrightarrow ~~\{\psi\simeq 0\},
	\ee
where the symbol $\simeq$ means 
``cohomologically equivalent ". Therefore, if we want to complete the proof of the correspondence between 
the
$|\mbox{phys}\rangle$ states of (101) and the
$\psi$ of (102), we must show that:
	\be
	\label{alfa}
	\{|\mbox{phys}\rangle = Q_{\scriptscriptstyle (1)} |\chi\rangle
	~~~\mbox{with}~~~\widetilde{\mathcal H}
	|\chi\rangle = 0\}~~\Longrightarrow ~~\{|\mbox{phys}\rangle \simeq 0\}.
	\ee
\noindent 
Note that the state $|\chi\rangle$ is not 
required to be physical, but only to satisfy the LHS of eq.~(107); this implies
that $|\chi\rangle$ 
in general does not satisfy the third and the fourth conditions of (101). This
means that 
$|\chi\rangle$ in general can depend on $g$ and $\alpha$. The point is that this
dependence, 
due to the LHS of eq.~(\ref{alfa}) and to the requirement that $|\mbox{phys}\rangle$ does not
depend on $g$ 
and $\alpha$, must have the following form, as proven in Appendix F:
	\begin{equation}
	\label{beta}
	|\chi\rangle = |\chi_{0}\rangle + |\zeta;\alpha,g\rangle  
	\end{equation}
\noindent where $|\chi_{0}\rangle$ is independent of both $g$ and $\alpha$,
and 
$|\zeta;\alpha,g\rangle \in \ker Q_{\scriptscriptstyle (1)}$. Moreover 
if $|\chi\rangle\in\ker\widetilde{\mathcal H}$ (as imposed by eq.~(\ref{alfa})), also 
$|\chi_{0}\rangle\in\ker\widetilde{\mathcal H}$ as one can check by applying 
$ Q_{\scriptscriptstyle (1)}^2 = -i\widetilde{\mathcal H}$ to both members of 
eq.~(\ref{beta}). We are now ready to show that states of the form (\ref{alfa}) are
cohomologically 
equivalent to zero according to the cohomology defined by
$\Omega^{\scriptscriptstyle (eq.)}_{BRS}$ 
of eq.~(98). The proof goes as follows:
\be
\label{gamma}
\begin{array}{rl}
\mid\mbox{phys}\rangle & =Q_{\s (1)}\mid\chi\rangle \\
& = Q_{\s (1)}\mid\chi_{0}\rangle \\
& = [C^{\s (1)}]^{-1}[C^{\s (1)}Q_{\s (1)}+C^{\s H}\HT+{\mathcal P}_{\s (1)}
\Pi_{\s\alpha}+{\mathcal P}_{\s H}\Pi_{\s g}+i (C^{\s (1)})^{2}
{\o{\mathcal P}}_{\s H}]\mid\chi_{0}\rangle \\
& = [C^{\s (1)}]^{-1}\Omega^{\s eq.}_{\s BRS}\mid\chi_{0}\rangle \\
& = \Omega^{\s (eq.)}_{\s BRS}\mid\chi'\rangle 
\end{array}
\ee
\noindent (where we have defined $\mid\chi'\rangle\equiv[C^{\s (1)}]^{-1}
\mid\chi_{0}\rangle$) and therefore $\mid\mbox{phys}\rangle$ is cohomologically
equivalent to zero. 
Note that the ghost $ C^{\s (1)}$ is bosonic in character and so we can 
build its inverse. In the third equality of (\ref{gamma}) we have used the fact that the
second, 
the third and the fourth terms give zero when applied to $|\chi_{0}\rangle$. 
The last term ($i (C^{\s (1)})^{2}
{\o{\mathcal P}}_{\s H}$) also annihilates $|\chi_{0}\rangle$ by a similar
reasoning based on the 
fact that $\mid\mbox{phys}\rangle$ cannot contain any dependence on $C^{\s H}$.
This concludes our proof.

Basically with our path-integral we have managed
to get the propagation of equivariantly non-trivial states by properly gauging
the susy. This is what the susy is telling us from a
geometrical point of view. It is not the first time that the equivariant
cohomology is
reduced to a sort of BRS formalism\scite{Niem}, but differently from these 
authors the BRS-BFV charge we obtained  is really linked to a local 
invariance problem associated to the lagrangian (\ref{eq:equi-sei}).
We feel that this detailed analysis of the problem will be appreciated
by the community of physicists not last for his pedagogical value.

At this point we can say that, by using the trick of gauging it,
we have understood the geometry lying behind our susy.
Most probably a further geometrical understanding of our 
original path-integral (\ref{eq:sei})
will emerge once we 
gauge\scite{Hilb} also some of the other universal symmetries found
previously\scite{Ennio}\scite{DFG}. These are made of the generators
of an Isp(2) superalgebra and of a noncanonical charge\scite{DFG}
whose role is to rescale the entire action of any classical system.
To better understand the geometry lying behind them we plan to study 
the effect of these local symmetries on the superspace which is 
made\scite{Ennio}\scite{DFG} of the time $t$ and of two Grassmannian 
partners of the time which we called $\theta,{\o\theta}$.

We feel that this attempt to better understand the geometry lying behind
our functional formulation of CM will help not only in unveiling 
some of the  unsolved problems of Classical Mechanics 
(ergodicity and integrability being some of them) but also 
in throwing light  on the transition to the
quantum regime which, one-hundred years after the invention of
$\hbar$, is still an issue in need of a deeper undestanding.
The reader may wonder what the quantization has to do with 
the geometry lying behind our classical path-integral and its associated
superspace
($t,\theta,{\bar\theta}$). The explanation can be found
in ref.\scite{Planc} where it was shown that the process of quantization
is equivalent to freezing to zero the Grassmannian partners of
time $(\theta,{\bar\theta})$. This freezing to zero must be 
hiding some deep geometrical structures which we have not 
completely understood yet. The same may happen for the new universal
symmetry that we discovered in ref.\scite{DFG}. This is a symmetry
which rescales the entire action of any classical system and it is
clearly broken at the quantum level by the presence of $\hbar$.
We will return to these problems in other papers.

In this paper we will continue to concentrate on Classical Mechanics
and try to put the geometrical basis to attack in our formalism
issues like integrability or ergodicity. This is what we
plan to do in the next section.

\section{Motion On Constant-Energy Surfaces.}
\indent
In the previous sections we have gauged the susy and we 
have ended up with a constrained motion on the hypersurfaces
given by eq.(\ref{eq:sessantuno}) or (\ref{eq:equi-otto}). 
In this section we shall reverse the procedure. We want to constrain 
the motion on some particular hypersurfaces of phase-space
and see which local symmetries the associated lagrangian will
exhibit. The hypersurfaces we
choose are those defined by fixed values of the constants of 
motion. We will explain later the reasons for this choice.

Let us start with the constant energy surface: $H(p,q)=E$.
The most natural thing to do is to add this constraint to the $\LT$ of
eq.(\ref{eq:sette}):

\be
\label{eq:energia-uno}
\LT_{\s E}\equiv\LT+f(t)(H-E)
\ee

\noindent where $f(t)$ is a gauge variable.
This lagrangian has the following local invariance :

\be
\label{eq:energia-due}
\left\{
\begin{array}{l}
\delta_{\s H}(\cdot)  =  [\tau(t) H,(\cdot)]\\
\delta_{\s H}f(t)  =  i{\dot\tau}(t)
\end{array}
\right.
\ee

\noindent
where $\tau(t)$ is an infinitesimal bosonic parameter
and we have indicated
with $(\cdot)$ any of the variables \quattrova.
Anyhow this is not the whole story. In fact if we restrict the original 
phase-space to be a constant  energy surface, 
the forms $c^{a}$ themselves must be restricted to be
those living only on the energy surface, that means they must be 
``perpendicular" to the gradient of the Hamiltonian. 
This constraint is:

\be
\label{eq:energia-tre}
c^{a}\partial_{a}H=0=N_{\s H}
\ee

\noindent
So basically we have to impose that the $N_{\s H}$-function of eq.
(\ref{eq:trentanove})
be zero. This is a further constraint we should add to the $\LT_{\s E}$ of
eq.(\ref{eq:energia-uno}).  One may think that an analogous restriction has 
to be done also for the vector fields considering that  forms and tensor
fields are paired  by the symplectic matrix as explained
in eqs.(\ref{eq:trentadue})(\ref{eq:trentatre}). If we accept
this we will have to add  the condition that the
${\o N}_{\s H}$ of eq.(\ref{eq:trentanove}) be zero.

A manner to get all these constraints automatically, without
having to add them by hand, is beautifully achieved if we
request that the new Hamiltonian $\HT_{\s E}$, which describes the motion 
on the energy surface, be a Lie-derivative of an Hamiltonian flow
like the original $\HT$ was. $\HT_{\s E}$ must be a Lie-derivative because,
after all, the motion is the same as before. This time the difference is 
that we fix a particular value of the energy and so we include this initial
condition directly into the lagrangian. If $\HT_{\s E}$ is a Lie-derivative
of an Hamiltonian flow then, from what we said below
eq.({\ref{eq:trentotto}), we gather that $\HT_{\s E}$ must be of
the following form:

\be
\label{eq:energia-quattro}
\HT_{\s E}=[\Qb^{\s E}[\QBb^{\s E},(\cdot)]]
\ee

\noindent
that means it must be the BRS variation of the antiBRS variation
of some function that we  indicate in (\ref{eq:energia-quattro})
 with $(\cdot)$. The BRS and antiBRS
charges in (\ref{eq:energia-quattro}) are not the ones 
of $\HT$, that means those of eqs.(\ref{eq:venti})(\ref{eq:ventuno}). For
this reason 
we have indicated them with  different symbols. 

If  $\HT_{\s E}$ is of the form above then the associated lagrangian must be
BRS invariant. Let us start from the $\LT_{\s E}$ in eq.(\ref{eq:energia-uno})
and see if it is BRS invariant at least under the old $\Qb$. It is easy to
do that calculation and we get:

\be
\label{eq:energia-cinque}
[\Qb,\LT+f(H-E)]=f(t)N_{\s H}\neq 0
\ee

\noindent
So it is not BRS invariant. The way out 
is to add to $\LT_{\s E}$ the $N_{\s H}$ multiplied by a gauge field.
 The new lagrangian
is:

\be
\label{eq:energia-sei}
\LT^{\prime}_{\s E}\equiv\LT+f(t)(H-E)+i\alpha(t) N_{\s H}
\ee

\noindent
where $\alpha(t)$ is the  Grassmannian gauge field. This lagrangian is
BRS-invariant
provided that we define a proper BRS-variation also on the gauge-fields
$\alpha(t)$ 
and $f(t)$. These proper BRS transformations  are:

\be
\label{eq:energia-sette}
\left\{
\begin{array}{l}
\delta (\cdot)  =  [\e \Qb,(\cdot)] \\
\delta {\alpha}  =  i\e f \\
\delta f=0
\end{array}
\right.
\ee

\noindent
where we have indicated with $(\cdot)$ any of the \quattrova and with $\Qb$ the
old
BRS charge of eq.(\ref{eq:venti}).

Next let us notice that if the form of $\HT_{\s E}$ is the one of eq.
(\ref{eq:energia-quattro}) 
then the associated
lagrangian has to be also antiBRS-invariant. Let us check if this happens with
the
$\LT^{\prime}_{\s E}$ of eq.(\ref{eq:energia-sei}):

\be
\label{eq:energia-otto}
[\QBb,\LT^{\prime}_{\s E}]=f(t){\o N}_{\s H}-\alpha(t)\HT
\ee

\noindent
So it is not antiBRS invariant and  the way out is again to add 
to $\LT^{\prime}_{\s E}$
the generators appearing on the RHS of (\ref{eq:energia-otto}).
The final lagrangian is:

\be
\label{eq:energia-nove}
\LT^{\prime\prime}_{\s E}\equiv\LT+f(t)(H-E)+i\alpha(t)N_{\s H}+
i{\o \alpha}(t){\o N}_{\s H}-g(t)\HT
\ee

\noindent
where $(f,\alpha,{\o\alpha},g)$ are gauge-fields. So we see that the
request that our $\HT_{\s E}$ be a Lie-derivative (\ref{eq:energia-quattro})
has automatically
produced the constraints $N_{\s H}=0$ and ${\o N}_{\s H}=0$ that otherwise we
would
have had to add by hand like we did in the reasoning leading to
eq.(\ref{eq:energia-tre}).

The lagrangian $\LT^{\prime\prime}_{\s E}$ is invariant under the 
following generalized BRS and antiBRS transformations:

\be
\label{eq:energia-dieci}
\begin{array}{ll}
	\delta_{\s \Qb}\equiv
	\left\{
	\begin{array}{l}
	\delta(\cdot) = [\e\Qb,(\cdot)]\\
	\delta \alpha = i\e f\\
	\delta {\o\alpha} = 0\\
	\delta f = 0\\
	\delta g =\e{\o\alpha}\\
	\end{array}
	\right. 
	&
	{\o{\delta}}_{\s\QBb}\equiv
	\left\{
	\begin{array}{l}
	{\o\delta}(\cdot) = [{\o\e}\QBb, (\cdot)]\\
	{\o\delta}\alpha  = 0\\
	{\o\delta}{\o\alpha}  = i{\o\e} f\\
	{\o\delta}f = 0\\
	{\o\delta}g=-{\o\e}{\alpha}
	\end{array}
	\right. 

\end{array}
\ee

\noindent
It is straightforward to build the BRS-antiBRS charges which produce the
variations indicated above. They are:

\bea
\label{eq:energia-dodici-tredici}
\Qb^{\s E}\equiv \Qb + i f\Pi_{\s \alpha}+i{\o\alpha}\Pi_{\s g} \\
\QBb^{\s E}\equiv \QBb + i f \Pi_{\s{\o\alpha}}-i\alpha\Pi_{\s g}
\eea

\noindent
where $\Pi_{\s\alpha}$, $\Pi_{\s{\o\alpha}}$, $\Pi_{\s g}$ are the
momenta conjugate to the variables $\alpha,{\o\alpha}, g$ and their
graded commutators are:

\be
\label{eq:energia-undici}
[\alpha,\Pi_{\s\alpha}] = [{\o\alpha},\Pi_{\s{\o\alpha}}]= i[\Pi_{\s g}, g ]
 = i [f, \Pi_{\s f}]=1
\ee

\noindent

The new BRS and antiBRS charges are nilpotent, as BRS charges should be, 
and anticommute among themselves

\be
\label{eq:energia-quattordici}
({\Qb^{\s E}})^{2}=({\QBb^{\s E}})^{2}=[\Qb^{\s E},\QBb^{\s E}]=0
\ee

\noindent
Having obtained these charges it is then easy to prove that the 
$\HT_{\s E}^{\prime\prime}$ associated to the $\LT^{\prime\prime}_{\s E}$
of eq.(\ref{eq:energia-nove}) has the form (\ref{eq:energia-quattro})
with the $(\cdot)$ in (\ref{eq:energia-nove}) given by  $-i(H+g(H-E))$, 
i.e:
 
\be
\label{eq:energia-quindici}
\HT_{\s E}^{\prime\prime}=-i[\Qb^{\s E}[\QBb^{\s E},H+g(H-E)]
\ee

\noindent
This shows, with respect to the $\HT$ of eq.(\ref{eq:trentotto}), 
that the 0-form out of which the Hamiltonian vector field is built
is not $H$ but $H+g(H-E)$. This is natural in the sense that this
0-form feels the constraint $H-E=0$.

\noindent
The symplectic structure behind our construction
can be made more manifest if we introduce the following notation:

\be
\label{eq:energia-sedici}
	\left\{
	\begin{array}{l}
	\varphi^{\s A}\equiv (\varphi^{a}; g, \Pi_{\s f}) \\
	\lambda_{\s A}\equiv (\lambda_{a};\Pi_{\s g}, f)\\
	c^{\s A}\equiv(c^{a};{\o\alpha}, \Pi_{\s \alpha})\\
	{\o c}_{\s A}\equiv ({\o c}_{a};\Pi_{\s {\o\alpha}},\alpha) 
	\end{array}
	\right.
\ee

\noindent
where the index in capital letter $(\cdot)^{\s A}$ runs from 1 to $2n+2$ while
the one in small letters $(\cdot)^{a}$ runs from 1 to $2n$ and it refers to
the usual variables \quattrova. Let us also  introduce an enlarged symplectic 
matrix:

\begin{equation}
\label{eq:energia-diciassette}
	\omega^{\s AB}=
	\left(
	\begin{array}{cc}
	\omega^{ab} & 0 \\
	0 &     \left(
		\begin{array}{cc}
		0 & 1 \\
		-1 & 0
		\end{array}
		\right)
	\end{array}
	\right)
\end{equation}

\noindent
and then, using the definitions (\ref{eq:energia-sedici}) and 
(\ref{eq:energia-diciassette}),
the BRS-antiBRS charges (\ref{eq:energia-dodici-tredici})(121)
can be written in the following compact form:

\be
\label{eq:energia-diciotto}
	\left\{
	\begin{array}{l}
	\Qb^{\s E}=ic^{\s A}\lambda_{\s A} \\
	\QBb^{\s E}=i{\o c}_{\s A}\omega^{\s AB}\lambda_{\s B}.
	\end{array}
	\right.
\ee

\noindent
Note that this form resembles very much the one of the original BRS and antiBRS
charges
(\ref{eq:venti})(\ref{eq:ventuno}). It is also straightforward to prove that the
$\HT_{\s
E}^{\prime\prime}$ has an N=2
supersymmetry like the old $\HT$. To build the susy charges
we should first
construct the $(N_{H},{\o N}_{H})$ charges analogous to those in
(\ref{eq:trentanove}).
Replacing in (\ref{eq:trentanove}) the symplectic matrix and 
the variables with those constructed respectively in (\ref{eq:energia-diciassette})
and (\ref{eq:energia-sedici}), and the 0-form $H$ with the 
0-form $H+g(H-E)$ entering the $\HT_{\s E}$, we get:

\be
\label{eq:energia-diciannove}
\left\{
\begin{array}{l}
N^{\s E}_{\s H}= c^{\s A}\partial_{\s A}(H+g(H-E)) \\
{\o N}^{\s E}_{\s H}={\o c}_{\s A}\omega^{\s AB}\partial_{\s B}(H+g(H-E))
\end{array}
\right.
\ee
The supersymmetry charges analogous to those in eq.(\ref{eq:quaranta}) 
are then

\be
\label{eq:energia-venti}
	\left\{
	\begin{array}{l}
	Q_{\s H}^{\s E}\equiv \Qb^{\s E}-\beta N_{\s H}^{\s E} \\
	\QBH^{\s E}\equiv {\overline Q}^{\s E}_{BRS}+\beta{\o N}_{\s H}^{\s E}
	\end{array}
	\right.
\ee

\noindent
where $\beta$ is a dimensional parameter like the one appearing in
(\ref{eq:quaranta}). It is then easy to check  that:

\be
\label{eq:energia-ventuno}
[Q^{\s E}_{\s H}, \QBH^{\s E}]=2i\beta\HT_{\s E}^{\prime\prime}
\ee

\noindent Up to now we have found which are the global symmetries of our
lagrangian (\ref{eq:energia-nove}), but let us not forget that the
goal of this section was to find out if, by imposing a constraint
from outside like the one of being on a constant energy surface,
we would get a lagrangian with local symmetries. It is actually so
and a first hint was given by the local symmetry of 
eq.(\ref{eq:energia-due}). The full set of local invariances 
of the lagrangian $\LT_{\s E}^{\prime\prime}$ of eq.(\ref{eq:energia-nove})
is:

\be
\label{eq:energia-ventidue}
	\left\{
	\begin{array}{l}
	\delta(\cdot)=[\tau H+\bar\eta{\o N}_{\s H}+\eta N_{\s H}+\epsilon
	\HT,(\cdot)] \\
	\delta f=i{\dot \tau}\\
	\delta\alpha={\dot \eta}\\
	\delta{\bar\alpha}={\dot{\bar\eta}}\\
	\delta g=-i{\dot\epsilon} 
	\end{array}
	\right.
\ee
\noindent
where $(\tau,\eta,{\bar\eta},\epsilon)$ are the local gauge-parameters
depending on $t$, and with $(\cdot)$ we have indicated the variables
\quattrova.

The above local symmetry is not a local supersymmetry as in the previous
sections but a different graded one whose generators are $(H,N,{\o N},\HT)$.
While before, in section 3 and 4, the local symmetry was a clearly
recognizable one but the constraints --- being in the enlarged space
\quattrova --- were hard to visualize, here we have the inverse
situation: the constraint (the constant energy one) is easy to
visualize but not so much the local symmetries.

For a moment let us stop these formal considerations
and let us check that the Hamiltonian in eq. (\ref{eq:energia-quindici})
is the correct one.
The procedure we have followed here of constraining the motion
on a constant energy surface can be applied also to any other
constant of motion $I(\varphi)$. The result would be the following
Hamiltonian:

\be
\label{eq:energia-ventitre}
\HT^{\prime\prime}_{\s I}=\HT-f[I(\varphi)-k]-i\bar\alpha{\o N}_{\s (I)}-
i\alpha N_{\s (I)}+g{\widetilde{\cal I}}
\ee

\noindent where $k$ is a constant and

\be
\label{eq:energia-ventiquattro}
\left\{
\begin{array}{l}
N_{\s (I)}=c^{a}\partial_{a}I \\
{\o N}_{\s (I)}={\o c}_{a}\omega^{ab}\partial_{b}(I)\\
{\widetilde{\cal I}}=-i[\Qb,[\QBb, I(\varphi)]]
\end{array}
\right.
\ee

\noindent If we had an integrable system with $n$ constants of motion
$I_{\s i}$ in involution we would get as Hamiltonian the following one:

\be
\label{eq:energia-venticinque}
\HT^{\prime\prime}_{\s int.}=\HT-\sum_{i}\{f_{\s i}[I_{\s i}(\varphi)-k_{\s i}]
-i\bar\alpha_{\s i}{\o N}_{\s (I)_{\s i}}-
i\alpha_{\s i}N_{\s (I)_{\s i}}+g{\widetilde{\cal I}}_{\s i}\}
\ee

\noindent Let us now do a counting of the effective degrees of freedom of the
Hamiltonian $\HT^{\prime\prime}_{\s I}$ of eq.(\ref{eq:energia-ventitre}).
We have $8n$ variables \quattrova, plus 4 gauge fields 
$(f,\alpha,{\bar\alpha},g)$,
plus 4 momenta associated to these gauge fields minus 4 primary constraints
(which are the gauge-momenta equal zero), minus 4 secondary constraints
($I-k=0$, $N_{\s (I)}=0$, ${\o N}_{\s (I)}=0$, ${\widetilde{\cal I}}=0)$)
minus 8 gauge-fixings for a total of $8n-8$ independent phase-space variables.
For the Hamiltonian of an integrable system like
$\HT^{\prime\prime}_{\s int.}$ this counting would give $8n-8n=0$ as effective
number of phase-space variables describing the system. {\it This is absurd !}
This situation could already be seen in the one-dimensional harmonic oscillator
where $n=1$ and we have just one constant of motion (the energy). The 
number of variables of the associated $\HT^{\prime\prime}_{\s E}$ would be
$8n-8=8-8=0$. One could claim that our $\HT^{\prime\prime}_{\s int.}$,
 having zero degrees of
freedom, actually describes
a Topological-Field-Theory model, and maybe
it is so but for sure it does not describe the motion taking place on
the tori of an integrable system. On the tori we have the angles which
vary with time but here, having effectively zero phase-space variables,
we do not have any motion taking place at all. If it is a topological
theory  at most the 
$\HT^{\prime\prime}_{\s int.}$ can describe some static {\it geometric} feature
of our system. This in itself would be interesting and that is
why we have carried this construction so far.
We hope to come back to this issue in future papers but
for the moment we want to go back from where we started, that is
eq.(\ref{eq:energia-uno}) and see which is the way to get an Hamiltonian
describing really the motion on the constant energy surface.

What we basically want to get is an Hamiltonian whose counting of degrees
of freedom is correct. At the basic phase-space level labelled
by the variables $\varphi$ we have $2n$ variables minus 1 constraint
that is $H-E=0$ so the total number is $2n-1$. Going up to the
space \quattrova this number should be multiplied by 4 that is
$8n-4$. 

What went wrong in the construction of $\LT^{\prime\prime}_{\s E}$
of eq.(\ref{eq:energia-nove})? One thing that we requested, but  which 
was not necessary, was that the vector fields obey a constraint ${\o N}=0$
analogous to the one of the forms $N=0$. We made that request only in order 
to maintain the standard pairing between tensor fields and forms which
appear in any symplectic theory, but our theory is not a symplectic
one anymore because the basic space in $\varphi^{a}$ has odd dimension $2n-1$
and cannot be a symplectic space. So let us release the request of
having ${\o N}=0$. We could have a weaker request by adding this
constraint via the derivative of a lagrange multiplier (or gauge-field)
in the same manner  as we did in eq.(\ref{eq:ottantacinque}). There we realized
that adding constraints in this manner does not decrease the
number of degrees of freedom.
By consistency then  also the $\HT$ constraint,
which appeared together with the ${\o N}$ via 
the eq.(\ref{eq:energia-otto}), should be added via the derivative
of its associated lagrange multiplier. So in order to
describe the motion on constant energy surfaces, instead of
(\ref{eq:energia-nove}) the lagrangian we propose is:

\be
\label{eq:energia-ventisei}
L_{\s E}=\LT+f(H-E)+i\dot{\bar\alpha}{\o N}+i\alpha N-{\dot g}\HT
\ee\noindent
The constraints (primary and secondary) are:

\be
\label{eq:energia-ventisette}
\left\{
\begin{array}{l}
\Pi_{\s f}=0~;~~~~H-E=0~; \\
\Pi_{\s {\alpha}}=0~;~~~~N=0~;\\
\Pi_{\s\o\alpha}=-i{\o N}~; \\
\Pi_{\s g}=-\HT.
\end{array}
\right.
\ee

\noindent They are 6, all first class, and we need 6 gauge-fixings. So doing now
the counting of independent variables in phase-space
we have:  $8n$ variables \quattrova, plus $4+4$ gauge-fields and their momenta,
minus 6 constraints, minus 6 gauge fixings for a total of $8n-4$ which is
exactly the number we wanted!

Let us analyze the difference between the last constraint in
eq.(\ref{eq:energia-ventisette})\break 
(that is $\Pi_{\s g}=-\HT$) and the one
associated to the lagrangian $\LT^{\prime\prime}_{\s E}$ of eq.
(\ref{eq:energia-nove}) (that is $\HT=0$). This last constraint seems to
totally freeze the motion while the one in eq.(\ref{eq:energia-ventisette})
does not freeze it but just foliates the space of values of $\HT$.
Similar things can be said for the constraint ${\o N}=0$ associated
to $\LT^{\prime\prime}_{\s E}$ and the one, $\Pi_{\s \alpha}=i{\o N}$,
associated to  $L_{\s E}$. This last one would not force the vector
fields in a configuration symplectically equivalent to the one of forms.

Let us now proceed to further analyze the lagrangian $L_{\s E}$ 
of eq.(\ref{eq:energia-ventisei}).
The associated Hamiltonian is: 

\be
\label{eq:energia-ventotto}
H_{\s E}=\HT+\Pi_{\s\alpha}\dot{\alpha}+
\Pi_{\s g}{\dot g}+\Pi_{\s f}\dot{f}+\Pi_{\s\o\alpha}\dot{\bar\alpha}
-f(H-E)-i{\alpha}N-i\dot{\bar\alpha}{\o N}+\dot{g}\HT
\ee
\noindent
where we had to leave in some velocities because we could not perform
the Legendre transformation. From the above Hamiltonian we can go to
the canonical one\scite{Vinc} by imposing the 
primary constraints.
The result is:

\be
\label{eq:energia-ventinove}
H_{\s E}^{\s can.}\equiv\HT-f(H-E)-i{\alpha}N
\ee

\noindent
It is easy to prove that this $H_{\s E}^{\s can.}$ is a Lie-derivative
of a vector field but not of an Hamiltonian vector-field. To show that let
us first define the following new variables:

\be
\label{eq:energia-trenta}
\left\{
\begin{array}{l}
\varphi^{\s A}=(\varphi^{a},\pi_{\s f})\\
\lambda_{\s A}=(\lambda_{a},f)\\
c^{\s A}=(c^{a},\Pi_{\s{\alpha}})\\
{\o c}_{\s A}=({\o c}_{a},{\alpha})
\end{array}
\right.
\ee

\noindent
In this enlarged phase-space the BRS charge (or exterior derivative) is

\be
\label{eq:energia-trentuno}
Q_{\s BRS}^{\s can.}=\Qb+if\Pi_{\s{\alpha}}
\ee

\noindent
and the analog of the Hamiltonian vector field ${\o N}_{\s H}$ is

\be
\label{eq:energia-trentadue}
{\o N}_{\s H}^{\s can.}={\o N}_{H}-{\o\alpha}(H-E)
\ee

\noindent 
which is not an Hamiltonian vector field anymore because it cannot
be written as the antiBRS variation of something as a Hamiltonian
vector field should be (see eq.(\ref{eq:trentaquattro})).

The proof that $H_{\s E}^{\s can.}$ of eq.(\ref{eq:energia-ventinove})
is the Lie-derivative\scite{Marsd} of the vector field ${\o N}_{\s H}^{\s
can.}$ 
above comes from the fact that it can be written as the commutator 
of that vector field with the exterior derivative 
$Q_{\s BRS}^{\s can.}$  above:

\be
\label{eq:energia-trentatre}
H_{\s E}^{\s can.}=-i[Q_{\s BRS}^{\s can.},{\o N}_{\s H}^{\s can.}]
\ee

\noindent
To prove this relation  is straightforward. One just needs to
use the standard commutators plus the following ones:

\be
\label{eq:energia-trentaquattro}
[{\alpha},\Pi_{\s {\alpha}}]=1~~~;~~~[f,\Pi_{\s f}]=-i
\ee
\noindent
Eq.(\ref{eq:energia-trentatre}) implies that $H_{\s E}^{\s can.}$ of
eq.({\ref{eq:energia-ventinove})
is invariant under the global BRS transformations generated by
the $Q_{\s BRS}^{\s can.}$ of eq.(\ref{eq:energia-trentuno}). It is also easy to
see that the lagrangian $L_{\s E}$ of eq.(\ref{eq:energia-ventisei})
has the following local invariances different from those of
eq.(\ref{eq:energia-ventidue}):

\be
\label{eq:energia-trentaquattro}
	\left\{
	\begin{array}{l}
	\delta(\cdot)=[\tau H+{\bar\eta}{\o
	N}_{H}+{\eta}N_{H}+\epsilon
	\HT,(\cdot)] \\
	\delta f=i{\dot \tau}\\
	\delta\alpha=\dot{\eta}\\
	\delta{\bar\alpha}={\bar\eta}\\
	\delta g=-i{\epsilon} 
	\end{array}
	\right.
\ee

\noindent
Again, as before, this is a local symmetry but not a local supersymmetry.

Regarding the supersymmetry we can find a global one under which our 
$H_{\s E}^{\s can.}$ of eq.(\ref{eq:energia-ventinove}) is invariant.
It is the one generated by the following charge:

\be
\label{eq:energia-trentacinque}
Q_{susy}=Q_{\s BRS}^{\s can.}+{\o N}_{\s H}^{\s can.}
\ee

\noindent 
which is a susy charge because it is  easy to prove that:

\be
\label{eq:energia-trentasei}
[Q_{\s susy}]^{2}=i H_{\s E}^{\s can.}
\ee

\noindent
Differently from the $\HT$ of our original system, we do not have
an N=2 supersymmetry like in eq.(\ref{eq:equi-quattro}), but only an N=1
susy. This is due to the loss of a symplectic structure on
the constant energy surface.

The reason to work out this supersymmetry is not just academical. In fact we
proved in ref.\scite{Ergo} that there is a nice interplay between the
 loss of ergodicity of the system whose Hamiltonian
is $H$  and the spontaneous breaking of the susy 
of $\HT$. We proved in particular that if the susy of $\HT$ is unbroken 
then the system described by $H$ is in the ergodic phase and that if the system
is in the ordered phase (non-ergodic) then the susy of $\HT$ must be broken.
We could not prove the inverse of these two statements that is that if the
system is in the ergodic phase then the susy must be unbroken and that
if the susy is broken then the system must be in an ordered or non-ergodic
regime. The reason we could not prove these inverse statements was that
the energy at which the motion took place had not been specified.
We have no time here to explain the detailed reasons why this lack of
specification could not allow us to do the inverse of that statement
and we advice the reader interested in understanding this point
to study in detail the full set of papers contained in ref.\scite{Ergo}. 
The ergodicity\scite{Avez} is a concept
which is strongly energy dependent: a system can be ergodic at some energy
and not ergodic at other energies. So it was crucial to develop a formalism
giving us the motion on constant energy surfaces like we have done here.
The parameter $E$ entering our $H^{\s can.}_{\s E}$ is not a phase-space variable
and we can
consider it as a coupling constant. We know that  at some values of the 
coupling a symmetry can be broken while at others it can be restored. 
In  (\ref{eq:energia-ventisei}) the term containing the energy is like a 
tadpole term
because it is  proportional to a term linear in the field 
(the field in this case is
$f(t)$ while the coupling is $E$).

The attempt to have a formulation
of the CPI in which  $E$ enters explicitly was tried before\scite{Bill}
but along a different route. In that paper $E$ was not a coupling constant 
but a degree of freedom conjugate to time in a formulation of CM invariant 
under time-reparametrization. We think that, in order to understand the 
interplay {\it susy/ergodicity}, it is better to treat $E$ as a coupling constant. 

The next step would be to check whether the susy charge 
(\ref{eq:energia-trentacinque}) we have in $H_{\s E}^{\s can.}$
is that  for which the theorem\scite{Ergo}
mentioned above, regarding the interplay {\it susy/ergodicity}, holds also 
in the inverse form.
If this were the case  then we would have a criterion to check whether a system (at
some
energy) is ergodic or not using a universal symmetry like susy.
Maybe even a sort of Witten index could be built which, by signaling whether the susy
is broken or not, could tell us whether  the system is ergodic or not.

All this work will be left to future papers~\scite{Hilb}. Here we wanted to
stick to geometrical issues. We say ``future papers" because there 
are several other difficulties that have to be cleared before
really embarking on a full understanding of the interplay 
between susy and ergodicity. 
The main difficulty is  the presence of
zero and negative norm states which prevents the proper use of
something like a Witten index for the study of the above mentioned
interplay. 

\section{Conclusions}
In this paper we have continued the study of  the geometry lying
behind  a functional approach to CM developed in ref.\scite{Ennio}.
In particular here we have focused our attention on a universal supersymmetry
\scite{Ergo} which seemed to have a role both at the geometrical level,
for its relation to the issue of equivariant cohomology, and at the dinamical
level for its interplay with the concept of ergodicity\scite{Avez}.

We have clarified  the first connection by making the susy local
and building the BFV charge associated to this local symmetry.
This has been done in great detail in order to better understand
several issues present in the literature.
 
We have also put the geometrical basis to better understand the 
second connection, that is the interplay between susy and ergodicity. 
We have done it by formulating our system on constant energy surfaces and 
by thoroughly  exploring the geometry of this formulation.
We have brought to light  both the geometrical meaning
of the associated Hamiltonian  and also the surviving global and local 
symmetries. We hope now to have in our hands all the weapons needed for
the final attack on this issue of the interplay {\it susy/ergodicity}.

\newpage
\begin{center}
{\LARGE\bf Appendices}
\end{center}

\appendix
\makeatletter
\@addtoreset{equation}{section}
\makeatother
\renewcommand{\theequation}{\thesection.\arabic{equation}}

\section{Appendix }
\indent
In deriving eqs. (\ref{eq:quarantacinque}) and (\ref{eq:quarantasei}), or even
in checking
the global symmetry under $\QH$,
we had to work out things involving the variation of the kinetic
piece of $\LT$, i.e.:

\be
\label{eq:B-uno}
	[\e\QH,\l_{a}\dot{\p}^{a}+i\bc_{a}{\dot
	c}^{a}]=(\delta\l_{a})\dot{\p}^{a}+\l_{a}
	\frac{d}{dt}(\delta\p^{a})+i(\delta\bc_{a})\dot{c}^{a}+i\bc_{a}\frac{d}{dt}(\delta
	c^{a})
\ee

\noindent In this step we have interchanged the variation ``$\delta$" with
the time derivative ${d\over dt}$. If we actually do the time derivative
of a variation (for example of $\varphi^{a}$), we get

\be
\label{eq:B-due}
\begin{array}{ll}
	\begin{array}{rl}
	\displaystyle\frac{d}{dt}(\delta\p^{a}) & = \frac{d}{dt}[\e\QH,\p^{a}]
	\\
	& = [\frac{d}{dt}(\e\QH),\p^{a}] + [\e\QH,\frac{d\p^{a}}{dt}] \\
	& = [\e\QH,\frac{d\p^{a}}{dt}] \\
	& = \delta\left(\displaystyle\frac{d\p^{a}}{dt}\right),
	\end{array}
\end{array}
\ee

\noindent and if $\e$ is a global parameter the third equality in the equation
above holds (and as a result  we can interchange the variation
with the time derivative)  only if we use the fact that  ${d\QH\over dt}=0$. 
We have supposed the same thing in the case of the local variations
(\ref{eq:quarantacinque})(\ref{eq:quarantasei}), and the only
extra term appearing with respect to eq.(A.2) is the one containing
the ${\dot\epsilon}$.  
Using the conservation of $\QH$
means that we have assumed that the equations of motion hold.
Actually it is better not to assume that. In fact, if we
make this assumption, then the $\LT$ itself, of which we are checking the
invariance
via the variations above, would be zero. This is due to the fact that $\LT$ is
proportional
to the equations of motion and checking the invariance of
something that is zero is silly.  It is true that all our
path-integral does is to force us on the classical equations of motion,
but still it is better not to use that explicitly.

To avoid that problem the trick to use is to define the following
integrated charge:

\be
\label{eq:B-tre}
	\widetilde{\QH}=\int_{0}^{T}\QH(t)dt,
\ee

\noindent where $0$ and $T$ are the endpoints of the interval over which we
consider our motion.

It is then easy to check that all steps done in eq.(\ref{eq:B-due}),
once we replace $\QH$ with ${\widetilde\QH}$, can go 
 through without assuming the conservation of $\QH$. 
In fact  the ${d\widetilde\QH\over dt}$ in the third step of equation 
(\ref{eq:B-due})
is zero not because of the conservation of $\QH$ but because ${\widetilde\QH}$
is independent of $t$. Moreover the variation $\delta$ generated
by the ${\widetilde\QH}$ is the same as the one generated by $\QH$. This is so
because in checking the variation induced by ${\widetilde\QH}$
we have to use the non-equal-time commutators given by the path-integral
(\ref{eq:sei}) which are:

\begin{equation}
\label{eq:B-quattro}
[\phi^{a}(t),\lambda_{b}(t^{\prime})]=i\delta^{a}_{b}
\delta(t-t^{\prime})
\end{equation}

\noindent and similarly for the $c^{a}$ and ${\o c}_{a}$.

This charge was introduced before in the literature\scite{Blau} in order to
handle things in an abstract ``loop space". In our case we need 
that charge  for the much simpler reasons explained above.

                                                           
\section{Appendix }
\indent

\noindent In this appendix we will show what happens when we combine
two susy transformations.

Let us define the following two transformations $G_{\e_{\s 1}}$,
$G_{\e_{\s 2}}$ on any of the variables \quattrova (which we will collectively 
indicate with $O$).

\begin{equation}
\label{eq:C-uno}
	\begin{array}{l}
	\delta_{1}O\equiv[G_{\e_{\s 1}}, O]\equiv[\o\e_{1}\QBH + \e_{1}\QH, O]
	\\
	\delta_{2}O\equiv[G_{\e_{\s 2}}, O]\equiv[\o\e_{2}\QBH + \e_{2}\QH, O] 
	\end{array}
\end{equation}

\noindent where the infinitesimal parameters $\e_{\s 1}$,$\e_{\s 2}$ are time
dependent.

Combining two of these transformations we get

\begin{equation}
\label{eq:C-due}
	[\delta_{1},\delta_{2}]O=[G_{\e_{\s 1}},[G_{\e_{\s 2}},O]]-
	[G_{\e_{\s 2}},[G_{\e_{\s 1}},O]].
\end{equation}

\noindent Applying the Jacobi identity on the  RHS of (\ref{eq:C-due}), we get

\begin{equation}
\label{eq:C-tre}
	[\delta_{1},\delta_{2}]O=[[G_{\e_{\s 1}},G_{\e_{\s 2}}],O].
\end{equation}

\noindent By remembering eq. (\ref{eq:quarantuno})
it is  easy to work out what  $[G_{\e_{\s 1}}, G_{\e_{\s 2}}]$ is :
\begin{equation}
\label{eq:C-quattro}
	[G_{\e_{\s 1}},G_{\e_{\s
	2}}]=2i\beta(\o{\e}_{1}\e_{2}+\e_{1}\o{\e}_{2})\HT;
\end{equation}

\noindent inserting (\ref{eq:C-quattro}) in (\ref{eq:C-tre}), we get

\begin{equation}
\label{eq:C-cinque}
	\begin{array}{rl}
	[\delta_{1},\delta_{2}]O & 
	= 2i\beta(\o{\e}_{1}\e_{2}+\e_{1}\o{\e}_{2})[\HT,O] \\
	&= 2\beta(\o{\e}_{1}\e_{2}+\e_{1}\o{\e}_{2})\displaystyle\frac{dO}{dt}.
	\end{array}
\end{equation}

\noindent So we see from here that the composition of two local susy
transformations produces a local time-translation with parameter ${\o\e}_{\s 1}\e_{\s 2}+
\e_{\s 1}{\o\e}_{\s 2}$.

It is also instructive to do the composition of two {\it finite} 
susy transformations, the first ($G_{\s 1}$) with parameter $\e_{\s 1}$ 
and the other $(G_{\s 2}$) with parameter $\e_{\s 2}$.
The transformed variable $O^{\prime}$ has the expression:

\begin{equation}
\label{eq:C-sei}
	O'= \mbox{e}^{iG_{\s 1}}\mbox{e}^{iG_{\s 2}}\; O
	\;\mbox{e}^{-iG_{\s 2}}\mbox{e}^{-iG_{\s 1}}.
\end{equation}

\noindent Using the Baker-Hausdorff identity on the RHS above, we obtain:

\begin{equation}
\label{eq:C-sette}
	\begin{array}{rl}
	O' & = \mbox{e}^{[iG_{1} + iG_{2} -\frac{1}{2}[G_{1},G_{2}]]}\;
	O \; \mbox{e}^{[-iG_{1} - iG_{2} +\frac{1}{2}[G_{1},G_{2}]]} \\
	&= \mbox{e}^{i[\o{\e_{1}}\QBH + \e_{1}\QH + \o{\e_{2}}\QBH + \e_{2}\QH
	-\beta(\o{\e_{1}}{\e_{2}} + \e_{1}\o{\e_{2}})\HT]} \; O \;
	\mbox{e}^{-i[\o{\e_{1}}\QBH +
	\e_{1}\QH + \o{\e_{2}}\QBH + \e_{2}\QH -\beta(\o{\e_{1}}{\e_{2}} +
	\e_{1}\o{\e_{2}} )\HT]} \\
	&= \mbox{e}^{i\o{\gamma}\QBH + i\gamma\QH + i\Delta t \HT}\; O
	\;\mbox{e}^{-i\o{\gamma}\QBH-i\gamma\QH - i\Delta t\HT},
	\end{array}
\end{equation}

\noindent where $\o\gamma$, $\gamma$ and $\Delta t$ are respectively

\begin{equation}
\label{eq:C-8}
	\left\{
	\begin{array}{l}
	\gamma=\e_{1} + \e_{2} \\
	\o{\gamma} = \o{\e_{1}} + \o{\e_{2}} \\
	\Delta t = -\beta(\o{\e_{1}}\e_{2}+\e_{1}\o{\e_{2}}).
	\end{array}
	\right.
\end{equation}

\noindent So we see from eq.(\ref{eq:C-sette}) that the composition of two
finite local susy
is a local susy plus a local time-translation.

We will write down here how the variables \quattrova transform under a local
time
translation:

\be
\label{eq:C-nove}
\left\{
\begin{array}{l}
\delta^{\s
loc}_{\HT}\phi^{a}=[\eta(t)\HT,\phi^{a}]=-i\eta~\omega^{an}\partial_{n}\HT\\
\delta^{\s
loc}_{\HT}\lambda_{a}=[\eta(t)\HT,\lambda_{a}]=i\eta~\partial_{a}\HT\\
\delta^{\s
loc}_{\HT}c^{a}=[\eta(t)\HT,c^{a}]=-i\eta~\omega^{an}\partial_{n}\partial_{l}Hc^{l}\\
\delta^{\s loc}_{\HT}{\bar c}_{a}=[\eta(t)\HT,{\bar c}_{a}]=i\eta~{\bar
c}_{m}\omega^{mn}\partial_{n}\partial_{a}H.
\end{array}
\right.
\ee

\section{Appendix}
\indent
In this appendix we  analyze the question of how the transformations
(\ref{eq:cinquantacinque}) are generated by our first class constraints
(\ref{eq:cinquantasei})(\ref{eq:sessantuno}). This is a delicate issue
which is explained in detail on page 75 ff of ref.\scite{Teitel}. In fact naively
the transformation on $g$ contained in (\ref{eq:cinquantacinque}) apparently
cannot be obtained by doing the commutator of $g$ with the
proper gauge generators. The authors of ref.\scite{Teitel} are 
aware of similar problems and they 
suggested the following approach. First let us build an extended action
defined in the following way:
\be
\label{eq:D-uno}
S_{ext.}=\int dt [\LTloc +\pipsi{\dot \psi}+\pipsibar{\dot{\o\psi}}+
\pigi{\dot g}-U^{(i)}G_{i}]
\ee

\noindent
where the $G_{i}$ are all the six first class constraints
(\ref{eq:sessantacinque})
(and not just the primary ones)
and the $U^{(i)}$ the relative Lagrange multipliers. A general gauge
transformation
on an observable $O$ will be:

\be
\label{eq:D-due}
\delta O=[{\o\e} {\QBH}+\e \QH
+\eta\HT+{\o\alpha}\pipsi+{\alpha}\pipsibar+\beta\pigi,O]
\ee
\noindent
where $({\o\e},\e,\eta, {\o\alpha},{\alpha},\beta)$ are six infinitesimal gauge
parameters
associated to the six generators $G_{i}$.
If we consider the Lagrange multipliers $U^{i}$ as functions of the basic
variables,
then they will also change under the gauge transformation above. As we do not
know
the exact expression of the $U^{i}$ in terms of the basic variables, we will
formally indicate their gauge variation as $\delta U^{i}$. Using this notation
it is then a simple but long calculation to show that the gauge variation
of the action $S_{ext.}$ is:

\be
\label{eq:D-tre}
\begin{array}{rl}
\delta S_{ext}=\displaystyle\int dt & \left[
i{\dot\e}\QH-i{\dot{\o\e}}\QBH-i{\dot\eta}\HT-2i\HT
(\e\psi+{\o\e}{\o\psi})+\right. \\
&+{\o\alpha}\QBH+\alpha\QH-i\beta\HT+\pipsi{\dot{\o\alpha}}+\pipsibar
{\dot\alpha}-i\pigi{\dot\beta}+ \\
&-\delta U^{\s (2)}\pipsi-\delta U^{\s (1)}\pipsibar-\delta U^{\s(3)}\pigi-
\delta U^{\s (4)}\QH+ \\
&\left. -\delta U^{\s (5)}\QBH-\delta U^{\s(6)}\HT-U^{\s
(4)}{\o\e}2i\HT-U^{\s(5)}\e 2i\HT\right],
\end{array}
\ee

\noindent
where we have indicated  with $U^{\s(1)}$, for example, the 
Lagrange multiplier associated to the first of the constraints 
in (\ref{eq:sessantacinque}), with $U^{\s(2)}$ the one associated 
to the second and so on.

It is now easy to choose the variation of the Lagrange multipliers
in such a way to make $\delta S_{ext}=0$:

\be
\label{eq:D-quattro}
\begin{array}{ll}
\left\{
\begin{array}{l}
\delta U^{\s(1)}=-{\dot\alpha}\\
\delta U^{\s(2)}=-{\dot{\o\alpha}}\\
\delta U^{\s(3)}=-i{\dot{\beta}}
\end{array}
\right.
\;\;&;\;\;
\left\{
\begin{array}{l}
\delta U^{\s(4)}=\alpha-i{\dot\e}\\
\delta U^{\s(5)}={\o\alpha}-i{\dot{\o\e}}\\
\delta U^{\s(6)}=-i{\dot\eta}-i\beta-2i({\o\e}{\o\psi}+\e\psi)+2i({\o\e}
U^{\s(4)}+\e U^{\s(5)}).
\end{array}
\right.
\end{array}
\ee

\noindent
We can now  proceed as in ref.\scite{Teitel} by restricting the Lagrange
multipliers to be only those  of the primary fields (\ref{eq:cinquantasei})

\be
\label{eq:D-cinque}
U^{\s (4)}=U^{\s (5)}=U^{\s (6)}=0
\ee

\noindent
which implies that the gauge variations of these must be zero. From these two
conditions we get from eq.({\ref{eq:D-quattro}) the following
relations among the six gauge parameters

\be
\label{eq:D-sei}
\left\{
\begin{array}{l}
\alpha = i {\dot\e} \\
{\o\alpha} = i{\dot{\o\e}} \\
\beta = -{\dot\eta}-2({\o\e}{\o\psi}+\e\psi).
\end{array}
\right.
\ee
\noindent
As a consequence  the general gauge variation of an observable $O$
given in eq.(\ref{eq:D-due}) becomes

\be
\label {eq:D-sette}
\delta O =[{\o\e}\QBH+\e\QH+\eta\HT+i{\dot{\o\e}}\pipsi+
i{\dot\e}\pipsibar+(-{\dot\eta}-2({\o\e}{\o\psi}+\e\psi))\pigi,O]
\ee
\noindent
and applying it on the three variables $\gaug$
we get:

\be
\label{eq:D-otto}
\left\{
\begin{array}{l}
\delta\psi=[i{\dot{\o\e}}\pipsi,\psi]=i{\dot{\o\e}}\\
\delta{\o\psi}=[i{\dot\e}\pipsibar,{\o\psi}]=i{\dot\e}\\
\delta g =[-({\dot\eta}+2({\o\e}{\o\psi}+\e\psi)\pigi,g]=i{\dot\eta}+2i({\o\e}
{\o\psi}+\e\psi).
\end{array}
\right.
\ee

\noindent
This is exactly the transformation (\ref{eq:cinquantacinque}) obtained here
from the generators $G_{i}$ of eq.(\ref{eq:sessantacinque}).  This concludes
the explanation of how the variations (\ref{eq:cinquantacinque}), derived
from a pure lagrangian variation, could be obtained via the {\it canonical}
gauge generators $G_{i}$.
\section{Appendix}
\indent
In this appendix, for purely pedagogical reasons, we will show
how to gauge away the $\gaug$.

The infinitesimal transformations are given in eq.(\ref{eq:cinquantacinque})
and the first thing to do is to build  finite transformations out of the
infinitesimal ones. If we
start from a configuration $(\psi_{\scriptscriptstyle 0}(t),
{\o\psi}_{\scriptscriptstyle 0}(t),g_{\scriptscriptstyle 0}(t))$, after
one step we arrive at $(\psi_{\scriptscriptstyle 1}(t),
{\o\psi}_{\scriptscriptstyle 1}(t),g_{\scriptscriptstyle 1}(t))$
which are given by:
 
\begin{equation}
\label{eq:E-uno}
	\left\{
	\begin{array}{l}
	\psi_{1}(t)=\psi_{0}(t)+i\dot{\o{\e}}(t) \\
	\bpsi_{1}(t)=\bpsi_{0}(t)+i\dot{\e}(t)  \\
	g_{1}(t)=g_{0}(t)+i\dot{\eta}(t)+2i(\e(t)\psi_{0}(t)+{\o\e}(t)\bpsi_{0}(t)).
	\end{array}
	\right.
\end{equation}

\noindent
It is not difficult to work out what we get
after $N$ steps:
\begin{equation}
\label{eq:E-due}
	\left\{
	\begin{array}{l}
	\psi_{\s N}(t)=\psi_{0}(t)+iN\dot{\o{\e}}(t) \\
	\bpsi_{\s N}(t)=\bpsi_{0}(t)+iN\dot{\e}(t)  \\
	g_{N}(t)=g_{0}(t)+iN\dot{\eta}(t)+2i(N\e(t)\psi_{0}(t)+N{\o\e}(t)\bpsi_{0}(t))-
	N(N+1)(\e\dot{\o\e}+{\o\e}\dot{\e}).
	\end{array}
	\right.
\end{equation}

\noindent
Taking now the limit $N\rightarrow\infty$, but with the conditions:

\begin{equation}
\label{eq:E-tre}
	\left\{
	\begin{array}{l}
	N\e(t)\longrightarrow\Delta(t) \\
	N{\o\e}(t)\longrightarrow\o{\Delta}(t) \\
	N\eta(t)\longrightarrow\Delta_{g}(t) ,
	\end{array}
	\right.
\end{equation}

\noindent
where the various $\Delta(t)$ are non-divergent quantities, we get
that a finite transformation has the form:
\begin{equation}
\label{eq:E-quattro}
	\left\{
	\begin{array}{l}
	\psi(t)=\psi_{0}(t)+i\dot{\o{\Delta}}(t) \\
	\bpsi(t)=\bpsi_{0}(t)+i\dot{\Delta}(t)  \\
	g(t)=g_{0}(t)+i\dot{\Delta}_{g}(t)+2i(\Delta(t)\psi_{0}(t)+\o{\Delta}(t)\bpsi_{0}(t))-
	(\Delta\dot{\o\Delta}+{\o\Delta}\dot{\Delta}).
	\end{array}
	\right.
\end{equation}

\noindent
From the equation above it is easy to see that with the following
choice of $\Delta$'s

\begin{equation}
\label{eq:E-cinque}
	\left\{
	\begin{array}{l}
	\o{\Delta}(t)=i\displaystyle\int_{0}^{t}\psi_{0}(\tau)d\tau \\
	\Delta(t)=i\displaystyle\int_{0}^{t}\o{\psi}_{0}(\tau)d\tau   \\
	\Delta_{g}(t)=\displaystyle\int_{0}^{t}d\tau[ig_{0}(\tau)-2(\Delta\psi_{0}+{\o\Delta}
	\bpsi_{0})-
	i(\Delta\dot{\o\Delta}+{\o\Delta}\dot{\Delta})]
	\end{array}
	\right.
\end{equation}

\noindent
we can bring the $\gaug$ to zero.
We should anyhow be careful and check whether there are no obstruction
to this construction. Actually, after eq.(\ref{eq:quarantasei})
we said that, in order that the surface terms disappear,
we needed to require that $\e(t)$ and ${\o\e}(t)$ be zero at the
end-points $(0,T)$ of integration. From
eq.(\ref{eq:E-tre}) one sees that this implies:

\begin{equation}
\label{eq:E-sei}
	\begin{array}{l}
	\Delta(0)=\o{\Delta}(0)=0 \\
	\Delta(T)=\o{\Delta}(T)=0.
	\end{array}
\end{equation}

\noindent 
While the first condition is easily satisfied, as can be seen from
eq.(\ref{eq:E-cinque}), the second one would imply:

\begin{equation}
\label{eq:E-sette}
	\begin{array}{l}
	\Delta(T)=i\displaystyle\int_{0}^{T}\o{\psi}_{0}(\tau)d\tau = 0  \\
	\o{\Delta}(T)=i\displaystyle\int_{0}^{T}\psi_{0}(\tau)d\tau = 0.
	\end{array}
\end{equation}

\noindent
This is a condition which is not satisfied by any initial configuration
$\psi_{\scriptscriptstyle 0}$, ${\o\psi}_{\scriptscriptstyle 0}$ but only
by special ones. So we can say that, if we want transformations
which do not leave surface terms, then it may be impossible to gauge
away $(\psi,{\o\psi})$. Not to have surface terms may turn out to
be an important issue in some contexts. Anyhow this problem  does not
arise in the time-reparametrization transformation because, as we see
from eq.(\ref{eq:cinquantadue}), that transformation does not generate surface
terms.
\section{Appendix}
\indent
In this appendix we will show how the constraints ({\ref{eq:ottantasette}})
affect the Hilbert space of the system. We know that the {\it physical} states
should be annihilated by the constraints:

\be
\label{eq:F-uno}
\left\{
\begin{array}{l}
\,[\Pi_{\s\alpha}-{\Qb}]\mid\mbox{phys}\rangle=0\\
\,[\Pi_{\s{\o\alpha}}-{\QBb}]\mid\mbox{phys}\rangle=0
\end{array}
\right.
\ee
\noindent
and so this seems to restrict the original Hilbert space of the system.
On the other hand we have proved that the system obeying these constraints
and with lagrangian (\ref{eq:ottantacinque}) has the same number of degrees of
freedom as  the original system with lagrangian (\ref{eq:sette}) and moreover
they seem equivalent. If that is so then the Hilbert
space of the {\it physical} states should be equivalent or isomorphic to the
original Hilbert space. This is what we are going to prove in what follows.

Let us first solve the constraint (\ref{eq:F-uno}). The wave-functions
$\Psi(\cdots)$ of the system will depend not only on the $(\varphi^{a},c^{a})$
but also on the gauge-parameters $\alpha(t)$ and ${\o\alpha}(t)$. So equation
(\ref{eq:F-uno}) takes the form:

\be
\label{eq:F-due}
\left\{
\begin{array}{l}
\displaystyle\frac{\partial\Psi(\varphi,c;\alpha,{\o\alpha})}{\partial\alpha}=\Qb
\Psi(\varphi,c;\alpha,{\o\alpha})\\
\displaystyle\frac{\partial\Psi(\varphi,c;\alpha,{\o\alpha})}{\partial{\o\alpha}}=\QBb
\Psi(\varphi,c;\alpha,{\o\alpha})
\end{array}
\right.
\ee
\noindent
whose solution is

\be
\label{eq:F-tre}
\Psi(\varphi,c;\alpha,{\o\alpha})=\exp[\alpha\Qb+{\o\alpha}\QBb]~\psi(\varphi,c)
\ee
\noindent
where the $\psi(\varphi,c)$ are the states of the Hilbert space of the
old system  with lagrangian (\ref{eq:sette}) and the $\Qb$ and $\QBb$ 
should be interpreted as the differential operator associated
to the relative charge via the substitution (\ref{eq:diciassette}); the same
for all the Hamiltonians which we will use from now on. To prove that the
two systems are equivalent we should prove that there is an isomorphism
in Hilbert space between the solutions of the two Koopman-von Neumann\footnote{
By Koopman von Neumann equation we mean the analog of the Liouville
equations built via  the full $\HT$ and not via  just its bosonic part.}  
equations, the first one associated to the old Hamiltonian (\ref{eq:dodici}) 
and the second to the
Hamiltonian of the lagrangian (\ref{eq:ottantacinque}). This last one is
the {\it primary}\scite{Vinc} Hamiltonian:

\be
\label{eq:F-quattro}
\HT_{\s P}\equiv \HT_{can}+\mu(\Pi_{\s\alpha}-\Qb)+{\o\mu}
(\Pi_{\s{\o\alpha}}-\QBb)
\ee
where $\mu$ and ${\o\mu}$ are Lagrange multipliers and $\HT_{can}$
is the {\it canonical}\scite{Vinc} Hamiltonian associated to the Lagrangian
(\ref{eq:ottantacinque}).

The Koopman-von Neumann equation for this system is:

\be
\label{eq:F-cinque}
\HT {\s P}\Psi(\varphi,c,t;\alpha,{\o\alpha})=i
\frac{\partial\Psi(\varphi,c,t;\alpha,{\o\alpha})}{\partial t} 
\ee

\noindent which can be rewritten as:
\be
[\HT +\mu(\Pi_{\s\alpha}-\Qb)+{\o\mu}
(\Pi_{\s{\o\alpha}}-\QBb)] \Psi(\varphi,c,t;\alpha,{\o\alpha})= i
\frac{\partial\Psi(\varphi,c,t;\alpha,{\o\alpha})}{\partial t}.  
\ee
\noindent
Since $\Psi(\varphi,c,t;\alpha,{\o\alpha})$ is annihilated by the constraints
(E.1) we get:

\be
\label{eq:A}
\HT\Psi(\varphi,c,t;\alpha,{\o\alpha})=i
\frac{\partial\Psi(\varphi,c,t;\alpha,{\o\alpha})}{\partial t}; 
\ee

\noindent now we use (E.3) in (\ref{eq:A}) and this yields:

\be
\label{chicco}
\HT \exp [\alpha\Qb +{\o\alpha}\QBb]\psi(\varphi,c,t)= i
\frac{\partial}{\partial
t}\left[\exp(\alpha\Qb+{\o\alpha}\QBb)\psi(\varphi,c,t)\right].
\ee

\noindent The last step is to work out the derivatives in eq.(\ref{chicco}); we
obtain:

\be
\exp[\alpha\Qb+{\o\alpha}\QBb]~\HT\psi(\varphi,c,t) =  
\exp[\alpha\Qb+{\o\alpha}\QBb]~i\frac{\partial\psi(\varphi,c,t)}{\partial t}
\ee

\noindent which holds iff 

\be
\HT\psi=i\frac{\partial\psi}{\partial t}
\ee

\noindent and this concludes the proof that the two systems have not only the
same number of degrees of freedom but also the same Hilbert space.

\section{Appendix}
\noindent In this appendix we provide details regarding the derivation of eq.~(\ref{beta}).
Consider first the dependence on $g$. From 

\be
\mid\mbox{phys}\rangle = Q_{\scriptscriptstyle (1)} |\chi\rangle
\;\;\;\;\mbox{and}\;\;\;\;
\Pi_{\s g}\mid\mbox{phys}\rangle = 0
\ee 

\noindent we infer that 

\be
\Pi_{\s g}Q_{\s (1)} |\chi\rangle =  Q_{\s (1)}\Pi_{\s g}|\chi\rangle = 0,
\ee

\noindent which means that
\be
\label{ker}
\Pi_{\s g}|\chi\rangle \in \ker Q_{\scriptscriptstyle (1)}. 
\ee

\noindent If we represent $\Pi_{\s g}$ as $\Pi_{\s g}=-i\frac{\partial}{\partial
g}$,  eq.(\ref{ker}) implies:

\be
\displaystyle\frac{\partial}{\partial g}|\chi\rangle =
\sum_{m}f_{m}(g,\alpha)|\zeta_{m}\rangle
\ee

\noindent where $|\zeta_{m}\rangle$ form a basis of $\ker Q_{\s (1)}$. Solving
this last 
differential equation we get 

\be
\label{chi}
|\chi\rangle = |\chi_{0};\alpha\rangle + \sum_{m}\left[\int
dg~f_{m}(g,\alpha)\right]|\zeta_{m}\rangle
\ee

\noindent where $= |\chi_{0};\alpha\rangle$ does not depend on $g$ anymore. By
the same line of reasoning we 
can prove that $\Pi_{\s \alpha}|\chi\rangle \in \ker Q_{\scriptscriptstyle
(1)}$, which in turn implies that 
$\Pi_{\s \alpha}|\chi_0;\alpha\rangle \in \ker Q_{\scriptscriptstyle (1)}$. We
can repeat the 
previous steps and we arrive at the relation:

\be
|\chi_{0};\alpha\rangle = |\chi_{0}\rangle + \sum_{m}\left[\int d\alpha
~l_{m}(g,\alpha)\right]|\zeta_{m}
\rangle
\ee
    
\noindent which, substituted in eq.~(\ref{chi}), yields:

\be
|\chi\rangle = |\chi_{0}\rangle + \sum_{m}\left[\int d\alpha
~l_{m}(g,\alpha)\right]|\zeta_{m}\rangle + 
\sum_{m}\left[\int dg ~f_{m}(g,\alpha)\right]|\zeta_{m}\rangle \equiv
|\chi_{0}\rangle + 
|\zeta;\alpha,g\rangle,  
\ee

\noindent as we claimed in eq.~(\ref{beta}).

\section*{Acknowledgments}
We wish to thank W.D.Thacker and F.Legovini for collaboration
in the early stages of this project few years ago. Many thanks also 
D.Mauro for many helpful discussions and help.  
This work has been supported by grants from MURST and INFN of Italy.

\end{document}